\documentclass[twocolappendix,numberedappendix]{emulateapj}
\usepackage{natbib,amssymb,amsmath,graphicx,lscape,enumitem}
\renewcommand{\arraystretch}{1.1}

\newif\ifAMStwofonts
\AMStwofontstrue

\begin{document}

\title{COLD AND WARM ATOMIC GAS AROUND THE PERSEUS MOLECULAR CLOUD. II. \\
THE IMPACT OF HIGH OPTICAL DEPTH ON THE HI COLUMN DENSITY DISTRIBUTION \\
AND ITS IMPLICATION FOR THE HI-TO-H$_{2}$ TRANSITION}

\author{Min-Young Lee\altaffilmark{1}, 
Sne\v{z}ana Stanimirovi\'{c}\altaffilmark{2}, 
Claire E. Murray\altaffilmark{2}, 
Carl Heiles\altaffilmark{3}, 
Jesse Miller\altaffilmark{4}}
\altaffiltext{1}{Laboratoire AIM, CEA/IRFU/Service d'Astrophysique, Bat 709, 91191 Gif-sur-Yvette, France; 
min-young.lee@cea.fr}
\altaffiltext{2}{Department of Astronomy, University of Wisconsin, Madison, WI 53706, USA}
\altaffiltext{3}{Department of Astronomy, University of California, Berkeley, CA 94720, USA} 
\altaffiltext{4}{Department of Physics and Astronomy, Washington State University, PO Box 642814, Pullman, WA 99164-2814, USA}

\begin{abstract} 
We investigate the impact of high optical depth on the HI saturation observed in the Perseus molecular cloud 
by using Arecibo HI emission and absorption measurements toward 26 radio continuum sources. 
The spin temperature and optical depth of individual HI components are derived along each line-of-sight, 
enabling us to estimate the correction for high optical depth. 
We examine two different methods for the correction, Gaussian decomposition and isothermal methods,  
and find that they are consistent (maximum correction factor $\sim$ 1.2) 
likely due to the relatively low optical depth and insignificant contribution from the diffuse radio continuum emission for Perseus. 
We apply the correction to the optically thin HI column density on a pixel-by-pixel basis, 
and find that the total HI mass increases by $\sim$10\%. 
Using the corrected HI column density image and far-infrared data from the IRIS Survey, 
we then derive the H$_{2}$ column density on $\sim$0.4 pc scales. 
For five dark and star-forming sub-regions, 
the HI surface density is uniform with $\Sigma_{\rm HI}$ $\sim$ 7--9 M$_{\odot}$ pc$^{-2}$, 
in agreement with the minimum HI surface density required for shielding H$_{2}$ against photodissociation.
As a result, $\Sigma_{\rm H2}$/$\Sigma_{\rm HI}$ and $\Sigma_{\rm HI} + \Sigma_{\rm H2}$ show a tight relation. 
Our results are consistent with predictions for H$_{2}$ formation in steady state and chemical equilibrium, 
and suggest that H$_{2}$ formation is mainly responsible for the $\Sigma_{\rm HI}$ saturation in Perseus. 
We also compare the optically thick HI with the observed ``CO-dark'' gas, 
and find that the optically thick HI only accounts for $\sim$20\% of the ``CO-dark'' gas in Perseus.
\end{abstract} 


\section{Introduction}



As the most abundant molecular species in the universe, 
molecular hydrogen (H$_{2}$) is the main constituent of giant molecular clouds,  
the exclusive birthplaces of stars (e.g., \citeauthor{Kennicutt12} 2012). 
Recent observations of galaxies at both low and high redshifts have shown that 
star formation rates are strongly correlated with H$_{2}$ surface densities ($\Sigma_{\rm H2}$), 
the relation generally known as the ``Kennicutt--Schmidt law''
(e.g., \citeauthor{Schmidt59} 1959; \citeauthor{Kennicutt89} 1989; 
\citeauthor{Bigiel08} 2008; \citeauthor{Wilson09} 2009; \citeauthor{Tacconi10} 2010; 
\citeauthor{Schruba11} 2011; \citeauthor{Genzel13} 2013). 
This suggests that physical processes responsible for the atomic-to-molecular hydrogen (HI-to-H$_{2}$) transition
play a key role in the evolution of galaxies. 

Observationally, the HI-to-H$_{2}$ transition has been studied via 
ultraviolet (UV) absorption measurements along many random lines of sight through the Galaxy 
(e.g., \citeauthor{Savage77} 1977; \citeauthor{Rachford02} 2002; \citeauthor{Gillmon06a} 2006). 
For these measurements, either early-type stars or active galactic nuclei were used as background sources, 
and HI and H$_{2}$ column densities, $N$(HI) and $N$(H$_{2}$),   
were estimated from Lyman--alpha and Lyman--Werner (LW) band absorption. 
The UV studies probed the H$_{2}$ mass fraction $f_{\rm H2}$ = 2$N$(H$_{2}$)/[$N$(HI) + 2$N$(H$_{2}$)]
ranging from $\sim$10$^{-6}$ to $\sim$10$^{-1}$, 
and found that $f_{\rm H2}$ sharply increases at 
the total gas column density $N$(H) = $N$(HI) + 2$N$(H$_{2}$) of $\sim$(3--5) $\times$ 10$^{20}$ cm$^{-2}$. 
Additionally, the HI-to-H$_{2}$ transition has been indirectly inferred from the flattening 
of the relation between the HI column density and a tracer of total gas column density
(e.g., far-infrared (FIR) or hydroxide (OH) emission; 
\citeauthor{Reach94} 1994; \citeauthor{Meyerdierks96} 1996; \citeauthor{Douglas07} 2007; 
\citeauthor{Barriault10b} 2010; \citeauthor{Liszt14b} 2014). 
These studies found that the HI column density saturates to $\sim$(5--10) $\times$ 10$^{20}$ cm$^{-2}$, 
suggesting the presence of H$_{2}$. 
The HI saturation has also been found in extragalactic observations on $\sim$kpc scales
(e.g., \citeauthor{Wong02} 2002; \citeauthor{Blitz06} 2006; \citeauthor{Leroy08} 2008; \citeauthor{Wong09} 2009).  

Theoretically, the HI-to-H$_{2}$ transition has been investigated 
as a central process in photodissociation regions (PDRs). 
In PDRs, the interstellar medium (ISM) is predominatly atomic, 
and the molecular gas is only found in well-shielded regions
where dissociating UV photons are sufficiently attenuated. 
Many studies have been presented with different treatments of chemistry, geometry, and radiative transfer 
(e.g., \citeauthor{Spitzer48} 1948; \citeauthor{Gould63a} 1963; \citeauthor{Glassgold74} 1974; 
\citeauthor{vanDishoeck86} 1986; \citeauthor{Sternberg88} 1988; \citeauthor{Elmegreen93} 1993; 
\citeauthor{Draine96} 1996; \citeauthor{Spaans97} 1997; \citeauthor{Browning03} 2003; \citeauthor{Goldsmith07} 2007; 
\citeauthor{Liszt07} 2007; \citeauthor{Krumholz09} 2009; \citeauthor{Glover10} 2010; \citeauthor{Offner13} 2013; 
\citeauthor{Sternberg14} 2014), and an excellent summary of these studies was recently provided by \cite{Sternberg14}. 

Among the many studies, the \cite{Krumholz09} (KMT09 hereafter) model has recently been tested with  
a variety of Galactic and extragalactic observations
(e.g., \citeauthor{Bolatto11} 2011; \citeauthor{Lee12} 2012; \citeauthor{Welty12} 2012; 
\citeauthor{Wong13} 2013; \citeauthor{Motte14} 2014), 
thanks to its simple analytic predictions that allow a comparison with direct observables 
and an extrapolation of the model over a wide range of ISM environments.  
In the KMT09 model, a spherical cloud is illuminated by a uniform and isotropic radiation field, 
and the H$_{2}$ abundance is computed based on the balance between 
the rate of formation on dust grains and the rate of dissociation by UV photons (chemical equilibrium). 
The authors derived two dimensionless parameters that 
determine the location of the HI-to-H$_{2}$ transition in the cloud, 
resulting in the following important predictions.
First, they found that H$_{2}$ formation requires a certain amount of HI surface density ($\Sigma_{\rm HI}$) 
for shielding against the dissociating radiation field. 
Interestingly, this shielding surface density primarily depends on metallicity, 
and is expected to be $\sim$10 M$_{\odot}$ pc$^{-2}$ 
(corresponding to $N$(HI) $\sim$ 1.3 $\times$ 10$^{21}$ cm$^{-2}$) for solar metallicity.
Second, the H$_{2}$-to-HI ratio, $R_{\rm H2}$ = $\Sigma_{\rm H2}$/$\Sigma_{\rm HI}$, 
was predicted to linearly increase with the total gas surface density. 
This is because once the minimum HI surface density is obtained for shielding H$_{2}$ against photodissociation, 
all additional hydrogen is fully converted into H$_{2}$ and the HI surface density remains constant. 
As a result, $R_{\rm H2}$ is simply a function of metallicity and total gas surface density. 
Another interesting feature of the KMT09 model is that the ISM is ``self-regulated''
in that pressure balance between the cold neutral medium (CNM) and the warm neutral medium (WNM) 
determines the ratio of the UV intensity to the HI density. 

Aiming at testing the KMT09 model on sub-pc scales, 
we have recently focused on the Perseus molecular cloud (\citeauthor{Lee12} 2012).    
Perseus is one of the nearby molecular clouds in the Gould's Belt,
and is located at a distance of $\sim$300 pc (\citeauthor{Herbig83} 1983; \citeauthor{Cernis90} 1990). 
It has a projected angular size of $\sim$6$^{\circ}$ $\times$ 3$^{\circ}$ on the sky 
(based on the CO emission)\footnote{In this paper, $^{12}$CO($J = 1 \rightarrow 0$) is quoted as CO.},
and lies at high Galactic latitude $b$ $\sim$ $-$20$^{\circ}$, 
resulting in relatively simple HI spectra compared to other molecular clouds in the Galactic plane. 
With a total mass of $\sim$2 $\times$ 10$^{4}$ M$_{\odot}$ (\citeauthor{Sancisi74} 1974; \citeauthor{Lada10} 2010), 
Perseus is considered as a low-mass molecular cloud with an intermediate level of star formation (\citeauthor{Bally08} 2008). 
To test the KMT09 model, 
we derived $\Sigma_{\rm HI}$ and $\Sigma_{\rm H2}$ images 
using HI data from the Galactic Arecibo L-band Feed Array HI Survey 
(GALFA-HI; \citeauthor{Stanimirovic06} 2006; \citeauthor{Peek11} 2011) 
and FIR data from the Improved Reprocessing of the \textit{IRAS} Survey (IRIS; \citeauthor*{MD05} 2005). 
The final images were at $\sim$0.4 pc resolution, 
and covered the far outskirts of the cloud as well as the main body. 
We found that the HI surface density is relatively uniform with $\Sigma_{\rm HI}$ $\sim$ 6--8 M$_{\odot}$ pc$^{-2}$ 
for five dark and star-forming regions in Perseus (B5, B1E, B1, IC348, and NGC1333).
In addition, the relation between $R_{\rm H2}$ and $\Sigma_{\rm HI} + \Sigma_{\rm H2}$ on a log-linear scale 
was remarkably consistent for all individual regions, 
having a steep rise of $R_{\rm H2}$ at small $\Sigma_{\rm HI} + \Sigma_{\rm H2}$, 
a turnover at $R_{\rm H2}$ $\sim$ 1, and a slow increase toward larger $R_{\rm H2}$. 
All these results were in excellent agreement with the KMT09 predictions 
for solar metallicity\footnote{Perseus has solar metallicity (\citeauthor{GonzalezHernandez09} 2009).
See Section 7.2.1 of \citet{Lee12} for a detailed discussion.}, 
suggesting that the KMT09 model captures well the fundamental physics of H$_{2}$ formation on sub-pc scales. 

The observed HI saturation in Perseus, however, could alternatively result from the high optical depth HI.   
When the HI emission is optically thick, 
the brightness temperature ($T_{\rm B}$) becomes comparable to the kinetic temperature ($T_{\rm k}$). 
As a result, the HI surface density saturates in the optically thin approximation (which we used in \citeauthor{Lee12} 2012)
since $\Sigma_{\rm HI}$ $\propto$ $T_{\rm B}$ $\sim$ $T_{\rm k}$, 
and is underestimated. 
As the constant HI surface density is the key prediction from KMT09, 
it is critical to evaluate how much of the HI column density distribution is affected by the optically thick HI. 

In this paper, we assess the impact of high optical depth on the observed HI saturation in Perseus 
by using HI emission and absorption measurements obtained toward 26 background radio continuum sources 
(\citeauthor{Stanimirovic14} 2014; Paper I hereafter).  
These observations provide the most direct way to measure the high optical depth HI, 
allowing us to derive the ``true'' total HI column density distribution. 
Specifically, we use the HI emission and absorption spectra to estimate the correcton factor for high optical depth, 
and apply the correction to the HI column density image computed by \cite{Lee12} in the optically thin approximation. 
We take this localized approach rather than using data from existing all-sky surveys (e.g., \citeauthor{Heiles03a} 2003a; HT03a hereafter) 
in order to treat all spectra uniformly for the velocity range of Perseus 
and consider the possibility that CNM/WNM properties may vary with ISM environments 
(e.g., metallicity, star formation rate, etc.) as expected from theoretical models  
(e.g., \citeauthor{McKee77} 1977; \citeauthor{Koyama02} 2002; \citeauthor{Wolfire03} 2003; \citeauthor{Audit05} 2005; 
\citeauthor{MacLow05} 2005; \citeauthor{Kim13} 2013).

This paper is organized in the following way.  
We start with a summary of previous studies 
where various methods have been employed to derive the correction for high optical depth (Section \ref{s:background}). 
We then provide a description of the data used in this study (Section \ref{s:obs}). 
In Section \ref{s:correction}, we estimate the correction factor for high optical depth  
using two different methods, and compare our results with previous studies. 
In Sections \ref{s:perseus-corrected} and \ref{s:revisit-saturation}, 
we apply the correction to the HI column density image from \cite{Lee12} on a pixel-by-pixel basis, 
and revisit the HI saturation issue by rederiving the H$_{2}$ column density image 
and comparing our results with the KMT09 predictions. 
We then investigate whether or not the optically thick HI can explain the observed ``CO-dark'' gas in Perseus (Section \ref{s:CO-dark}),
and finally summarize our conclusions (Section \ref{s:summary}). 

\section{Background: Methods to estimate the correction for high optical depth} 
\label{s:background} 

In most radio observations,  
HI is detected in emission, 
and the intensity of radiation is measured as the brightness temperature as a function of radial velocity, i.e., $T_{\rm B}$$(v)$. 
Since the HI optical depth ($\tau$) can be measured 
only via absorption line measurements in the direction of background radio continuum sources, 
the optically thin approximation of $\tau \ll 1$ is frequently employed to estimate the HI column density: 
\begin{equation}
\label{eq:N_HI}
N(\textrm{HI})~(\textrm{cm}^{-2}) = 1.823 \times 10^{18} \int T_{\textrm{B}}(v)dv~(\textrm{K km s}^{-1}).
\end{equation}

Over the past three decades, 
several approaches have been employed to estimate 
how much of the true total HI column density is underestimated in the optically thin approximation. 
Most of these approaches can be classified as ``isothermal'', 
and the only multiphase approaches are by \citet{Dickey00} and HT03a. 
Here we summarize main results from some of the most important studies.

\cite{Dickey82} used 47 emission/absorption spectral line pairs in the direction of background sources, 
and estimated the ratio of the HI column density from the absorption spectra 
to the HI column density in the optically thin approximation. 
Although $\sim$1 at high and intermediate Galactic latitudes, 
the ratio reached $\sim$1.8 at low latitudes. 
There was considerable scatter in the ratio, however: 
several lines of sight at low latitudes showed small ratios, 
suggesting that the low latitude directions with large ratios likely intersect dense molecular clouds. 
In order to compute the ratios, HI in each velocity channel was assumed to have a single temperature (``isothermal'' approximation). 

In the \cite{Dickey00} study of the Small Magellanic Cloud (SMC), 
HI absorption observations were obtained in the direction of 13 background radio continuum sources.  
The corresponding emission spectra were derived  
by averaging HI profiles from \cite{Stanimirovic99} 
over a $3 \times 3$ pixel region (pixel size = 30$''$) centered on the position of each source.
The correction factor for high optical depth was calculated for each velocity channel in the isothermal approximation, 
and the line of sight integrated value $f$ was expressed as a function of the uncorrected $N$(HI): 
$f = 1 + 0.667(\log_{10}N(\textrm{HI}) - 21.4)$ for $N(\textrm{HI}) > 10^{21.4}$ cm$^{-2}$.
This relation was then applied to the $N$(HI) image of the SMC on a pixel-by-pixel basis, 
resulting in a $\sim$10\% increase of the total HI mass
from $\sim$3.8 $\times$ 10$^8$ M$_{\odot}$ to $\sim$4.2 $\times$ 10$^8$ M$_{\odot}$.
Although negligible at $N$(HI) $<$ $3 \times 10^{21}$ cm$^{-2}$,
the correction factor increased with the uncorrected $N$(HI)  
up to $\sim$1.4 at $N$(HI) $\sim$ 10$^{22}$ cm$^{-2}$.
In some cases, the correction factors for individual channels were larger than the integrated value, reaching up to $\sim$2. 
However, such values covered only a narrow range of channels, 
and their effect on $N$(HI) was relatively small. 
Finally, the authors rederived the correction factor in the two-phase approximation, 
and found that the difference between the one- and two-phase cases depends on 
the relative location of cold and warm HI components along a line of sight (Section \ref{s:method2} for details). 

In the Millennium Arecibo 21-cm Absorption Line Survey, 
HT03a obtained HI emission and absorption spectra toward 79 randomly positioned radio continuum sources,  
and performed Gaussian decomposition to estimate the physical properties 
of individual CNM and WNM components (column density, optical depth, spin temperature $T_{\rm s}$, etc.). 
These multiphase analyses showed that two or more components with very different spin temperatures
can contribute to a single velocity channel, 
implying that the isothermal treatment may not be satisfactory.
\citet{Heiles03b} (HT03b hereafter) then calculated the correction factor using the Gaussian decomposition results, 
which they called $R_{\rm raw}$ = $1/f$. 
There were interesting variations in $R_{\rm raw}$, 
ranging from $\sim$0.3 to $\sim$1.0 ($f$ = $\sim$1.0--3.0; Appendix B for details). 
Specifically, $f \sim 1.3$ was found for the Taurus/Perseus region.

A very different approach was adopted in \cite{Braun09} 
to calculate the correction for high optical depth in M31.  
They only used high-resolution HI emission observations for their modeling, 
and assumed that a single cold component determines the brightness temperature along a line of sight. 
While previous similar studies have applied a single temperature to the images of entire galaxies 
(e.g., \citeauthor{Henderson82} 1982; \citeauthor{Braun92} 1982), 
\cite{Braun09} estimated the spin temperature and non-thermal velocity dispersion for each pixel.  
After excluding HI spectra that likely suffer from 
a high blending of different components along a line of sight, 
they noticed that the opaque HI is organized into filamentary complexes and isolated clouds 
down to their resolution limit of $\sim$100 pc. 
The spin temperature was found to increase from $\sim$20 K to $\sim$60 K with radius to 12 kpc, 
and then to decline smoothly down to $\sim$20 K beyond 25 kpc. 
The estimated correction resulted in a $\sim$30\% increase of the global HI mass of M31.
Using the same methodology, \cite{Braun12} found that the correction for high optical depth 
increases the HI masses of the Large Magellanic Cloud (LMC) and M33 by the same amount ($\sim$30\%).
While the main advantage of the \cite{Braun09} approach is clearly that 
galactic-scale images of the opaque component can be produced solely from HI emission observations,
the method has several weaknesses. 
For example, it does not consider multiple components along a line sight 
and how they self-absorb each other. 
In addition, it does not take account of the possibility that 
some of the brightness temperature could come from unabsorbing warm HI components.

\begin{figure*}
\centering
\includegraphics[scale=0.08]{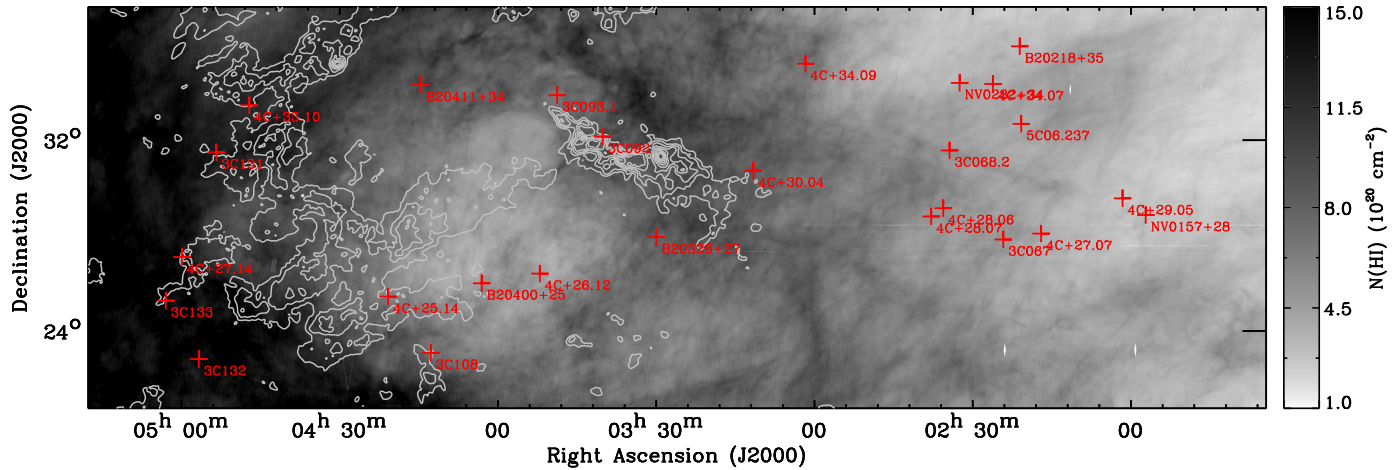}
\caption{\label{f:sources} 26 radio continuum sources overlaid on the HI column density image at 4$'$ resolution 
(4C$+$32.14 excluded; Section \ref{s:data-HI-abs} for details).  
The HI column density image is produced by integrating the GALFA-HI cube from
$v_{\rm LSR}$ = $-$5 km s$^{-1}$ to $+$15 km s$^{-1}$, 
and the gray contours are from the CfA CO integrated intensity image at 8.4$'$ resolution. 
The contour levels range from 10\% to 90\% of the peak value (69 K km s$^{-1}$) with 10\% steps.
In this figure, both the Perseus and Taurus molecular clouds are seen.}
\end{figure*}

\cite{Chengalur13} tested the optically thin and isothermal approximations 
with Monte Carlo simulations of the multiphase ISM.  
They varied the fraction of gas in three phases (CNM, WNM, and the thermally unstable neutral medium) 
and the location of each phase along a line of sight.  
A wide range of values were assumed for the HI column density (10$^{20}$--10$^{24}$ cm$^{-2}$) 
and the spin temperature (20--5000 K).
They found that the optically thin approximation underestimates the true HI column density 
by a factor of $\sim$1.6 when $\int \tau dv \sim 1$ km s$^{-1}$, 
while the underestimate can be as high as a factor of $\sim$20 when $\int \tau dv \sim 10$ km s$^{-1}$.
On the other hand, the simulations showed that the isothermal estimate tracks the true HI column density 
to better than 10\% even when $\int \tau dv \sim 5$ km s$^{-1}$. 
Their conclusion that the isothermal estimate provides a good measure of the true HI column density 
of up to $\sim$5 $\times$ 10$^{23}$ cm$^{-2}$ was insensitive 
to the assumed gas temperature distribution and the positions of the different phases along a line of sight.
We note that Equation (1) of \cite{Chengalur13} does not include the contribution 
from the cosmic microwave background (CMB) and the Galactic synchrotron emission, 
which can be significant in certain cases (e.g., low Galactic latitudes).
In addition, the authors did not consider self-absorption of the WNM by the foreground CNM. 

\cite{Liszt14b} compared $N$(HI) from Galactic HI surveys with $E(B-V)$ derived by \cite{Schlegel98}, 
and found a strong linear relation between $N$(HI) and $E(B-V)$ $\sim$ 0.02--0.08 mag 
and a flattening of the relation at $E(B-V) \gtrsim 0.08$ mag. 
While this flattening, likely due to H$_{2}$ formation, 
was essentially the same effect as what \cite{Lee12} found for individual regions in Perseus, 
the relation derived by \cite{Liszt14b} covers a large spatial area 
with randomly selected lines of sight predominantly at $|b| > 20^{\circ}$.
By using HI absorption data compiled by \cite{Liszt10}, 
the author then derived the correction for high optical depth, and applied it to the $N$(HI) data. 
The flattening at $E(B-V) \gtrsim 0.08$ mag persisted after the correction, 
confirming its origin in the onset of H$_{2}$ formation. 
The derived correction factor increased from $\sim$1.0 at $E(B-V)$ $\sim$ 0.01 mag 
to $\sim$1.4 at $E(B-V)$ $\sim$ 1 mag, 
and was $\lesssim$ 1.2 at $E(B-V) \lesssim 0.5$ mag. 

Recently, \cite{Fukui15} suggested a new approach 
to estimate the correction for high optical depth by using \textit{Planck} dust continuum data. 
They noticed that the dust optical depth at 353 GHz ($\tau_{353}$) correlates with the HI column density, 
and the dispersion in this relation becomes much smaller 
when the data points are segregated based on the dust temperature ($T_{\rm dust}$). 
The highest dust temperature was assumed to be associated with the optically thin HI, 
and the saturation seen in the $\tau_{353}$--$N$(HI) relation was then 
solely attributed to the optically thick HI.
By coupling the $\tau_{353}$--$N$(HI) relation with radiative transfer equations, 
\cite{Fukui15} calculated $T_{\rm s}$ and $\tau$ for the Galactic sky at $|b|$ $>$ 15$^{\circ}$ on a pixel-by-pixel basis. 
They found that more than 70\% of the data points have $T_{\rm s} < 40$ K and $\tau > 0.5$, 
and similar results were obtained for the high latitude molecular clouds MBM 53, 54, 55, and HLCG 92-35 \citep{Fukui14}.  
The correction for high optical depth resulted in a factor of $\sim$2 increase in the total HI mass in the solar neighborhood, 
implying that the optically thick HI may explain the ``CO-dark'' gas in the Galaxy.

\section{Data}
\label{s:obs}

\subsection{HI Emission and Absorption Observations} 
\label{s:data-HI-abs} 
We use the HI emission and absorption observations from Paper I. 
The observations were performed with the Arecibo telescope\footnote{
The Arecibo Observatory is operated by SRI International 
under a cooperative agreement with the National Science Foundation (AST-1100968),
and in alliance with Ana G. M\'{e}ndez-Universidad Metropolitana 
and the Universities Space Research Association.} using the L-band wide receiver, 
and were made toward 27 radio continuum sources located behind Perseus. 
The target sources were selected from the NRAO VLA Sky Survey (\citeauthor{Condon98} 1998) 
based on flux densities at 1.4 GHz greater than $\sim$0.8 Jy, 
and are distributed over a large area of $\sim$500 deg$^{2}$ centered on the cloud (Figure \ref{f:sources})\footnote{In this paper, 
we quote all velocities in the local standard of rest (LSR) frame, which is defined 
based on the average velocity of stars in the solar neighborhood: 
20 km s$^{-1}$ toward (R.A.,decl.) = (18$^{\rm h}$,30$^{\circ}$) in B1900.}. 
The angular resolution of the Arecibo telescope at 1.4 GHz is 3.5$'$.  
For the observations, a special procedure was adopted to make a ``17-point pattern'',
which includes 1 on-source measurement and 16 off-source measurements 
(HT03a; \citeauthor*{Stanimirovic05} 2005). 
This procedure was designed to consider HI intensity variations across the sky 
and instrumental effects involving telescope gains.
The data were processed using the reduction software developed by HT03a, 
and the final products for each source include 
an HI absorption spectrum ($e^{-\tau(v)}$), 
an ``expected'' HI emission spectrum ($T_{\rm exp}(v)$; 
HI profile that we would observe at the source position if the continuum source were not present), 
and their uncertainty profiles. 
Among the 27 sources, 4C$+$32.14 was excluded from further analyses 
because of its saturated absorption spectrum. 
With an average integration time of 1 hour, 
the root-mean-square (rms) noise level in the optical depth profiles was 
$\sim$1 $\times$ 10$^{-3}$ per 1 km s$^{-1}$ velocity channel. 
Finally, the derived optical depth and ``expected'' emission spectra 
were decomposed into separate CNM and WNM components using the technique of HT03a, 
and physical properties (optical depth, spin temperature, column density, etc.) 
were computed for the individual components. 
We refer to Paper I for details on the observations, data reduction, line fitting\footnote{Pros and cons of 
our Gaussian fitting method were discussed in HT03a in detail.
In the future, we plan to compare the Gaussian fitting method with results from numerical simulations 
to investigate biases that could be introduced by Gaussian fitting (Lindner et al. in prep).},
and CNM/WNM properties. 


\subsection{HI Emission Data from the GALFA-HI Survey}
\label{s:data-HI-ems}
In order to evaluate different methods for deriving the correction for high optical depth, 
we also use the HI emission data from the GALFA-HI survey (\citeauthor{Stanimirovic06} 2006; \citeauthor{Peek11} 2011). 
GALFA-HI uses ALFA, a seven-beam array of receivers at the focal plane of the Arecibo telescope, 
to map the HI emission in the Galaxy. 
Each of the seven dual polarization beams has an effective beam size of 3.9$'$ $\times$ 4.1$'$.

For Perseus, \citet{Lee12} produced an HI cube
centered at (R.A.,decl.) = (03$^{\rm h}$29$^{\rm m}$52$^{\rm s}$,$+$30$^{\circ}$34$'$1$''$) in J2000\footnote{In this paper, 
we quote all coordinates in J2000.} with a size of $\sim$15$^{\circ}$ $\times$ 9$^{\circ}$ 
by combining a number of individual GALFA-HI projects.
We use the same data here, but extend the HI cube up to $\sim$60$^{\circ}$ $\times$ 18$^{\circ}$ 
to include all radio continuum sources in Paper I. 
The HI column density image derived from the extended HI cube is shown in Figure 1 
along with our continuum sources (4C$+$32.14 excluded). 

\subsection{HI and H$_{2}$ Distributions of Perseus} 
\label{s:data-lee} 
We use the $N$(HI) and $N$(H$_{2}$) images from \citet{Lee12}. 
To derive the $N$(HI) image,  
we integrated the HI emission from $v_{\rm LSR}$ = $-$5 km s$^{-1}$ to $+$15 km s$^{-1}$ in the optically thin approximation. 
This velocity range was determined 
based on the maximum correlation between the $N$(HI) image and 2MASS-based $A_{V}$ data from the COMPLETE Survey \citep{Ridge06b}.   
In order to construct the $N$(H$_{2}$) image of Perseus with a large sky coverage,  
we used the 60 $\mu$m and 100 $\mu$m data from the IRIS survey (\citeauthor*{MD05} 2005), 
and derived the dust optical depth at 100 $\mu$m ($\tau_{\rm 100}$)
by assuming that dust grains are in thermal equilibrium. 
For this purpose, the emissivity spectral index of $\beta = 2$ was adopted, 
and the contribution from very small grains (VSGs) to the intensity at 60 $\mu$m ($I_{\rm 60}$) was removed 
by calibrating the derived $T_{\rm dust}$ image with DIRBE-based $T_{\rm dust}$ data from \citet{Schlegel98}. 
We then converted the $\tau_{\rm 100}$ image into the $A_{V}$ image 
by finding the conversion factor $X$ for $A_{V} = X \tau_{\rm 100}$ 
that results in the minimum difference between the derived $A_{V}$ and the COMPLETE $A_{V}$. 
This calibration of $\tau_{\rm 100}$ to the COMPLETE $A_{V}$ was motivated by \citet{Goodman09} 
who showed that dust extinction at near-infrared (NIR) wavelengths is the best probe of total gas column density. 
Finally, we measured a local dust-to-gas ratio (D/G)
by examining the $A_{V}$--$N$(HI) relation for diffuse regions, 
and derived the $N$(H$_{2}$) image by 
\begin{equation}
\label{eq:H2}
N({\rm H_{2}}) = \frac{1}{2}\left[\frac{A_{V}}{\rm D/G} - N({\rm HI})\right]. 
\end{equation}
The derived $N$(HI) and $N$(H$_{2}$) images are at 4.3$'$ resolution
(corresponding to $\sim$0.4 pc at the distance of 300 pc), 
and their median 1$\sigma$ uncertainties are 
$\sim$5.6 $\times$ 10$^{19}$ cm$^{-2}$ and $\sim$3.6 $\times$ 10$^{19}$ cm$^{-2}$. 
See Sections 3 and 4 of \citet{Lee12} for details 
on the derivation of the $N$(HI) and $N$(H$_{2}$) images and their uncertainties. 

\begin{figure}
\centering
\includegraphics[scale=0.53]{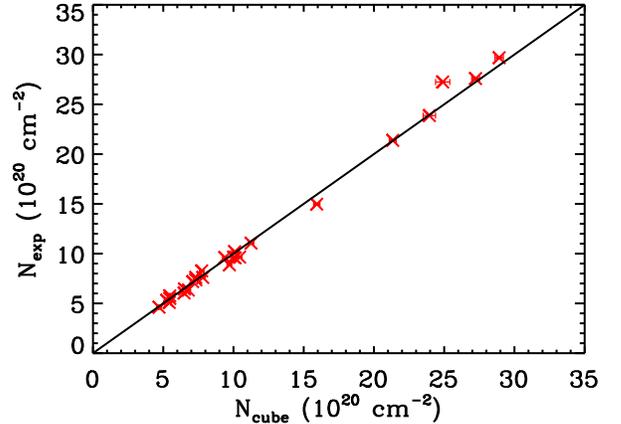}
\caption{\label{f:Ncube_Nexp} Comparison of the HI column densities calculated using the two different methods: 
$N_{\rm exp}$ from the spatial derivative method versus $N_{\rm cube}$ from the simple averaging method. 
Both quantities are estimated in the optically thin approximation, and the black solid line shows a one-to-one relation.} 
\end{figure}

\subsection{CO Data from the CfA Survey}
\label{s:data-cfa}
We use CO integrated intensity ($I_{\rm CO}$) data from \citet{Dame01}. 
\citet{Dame01} produced a composite CO survey of the Galaxy at 8.4$'$ resolution 
by combining individual observations of the Galactic plane and local molecular clouds. 
The observations were conducted with the Harvard-Smithsonian Center for Astrophysics (CfA) telescope. 
The final cube had a uniform rms noise of 0.25 K per 0.65 km s$^{-1}$ velocity channel. 
To estimate $I_{\rm CO}$ for Perseus, \citet{Dame01} integrated the CO emission 
from $v_{\rm LSR}$ = $-$15 km s$^{-1}$ to $+$15 km s$^{-1}$. 
See Section 2 of \citet{Dame01} for details on the observations, data reduction, and analyses.

\section{Correction for High Optical Depth}
\label{s:correction}
In the direction of 26 radio continuum sources,  
we measured the optical depth profiles which we use to estimate the true total HI column density, $N_{\rm tot}$. 
Along the same lines of sight, we also have the emission spectra 
that can be used to calculate the HI column density in the optically thin approximation, $N_{\rm low-\tau}$. 
This column density would be the only available information 
if no HI absorption data were present. 
In this section, we examine how $f = N_{\rm tot}/N_{\rm low-\tau}$, 
which we call the correction factor for high optical depth, varies with $N_{\rm low-\tau}$. 
Our aims are to offer an analytic estimate of $f(N_{\rm low-\tau})$ 
for Perseus using our HI emission and absorption measurements, 
and then to apply this correction to the $N$(HI) image from \cite{Lee12} on a pixel-by-pixel basis. 
In this way, we can account for the optically thick HI that was missed in the GALFA-HI emission observations. 
Since there are several approaches to derive $f(N_{\rm low-\tau})$, 
we first compare different methods using full line of sight information. 

\begin{figure*}
\centering
\includegraphics[scale=0.55]{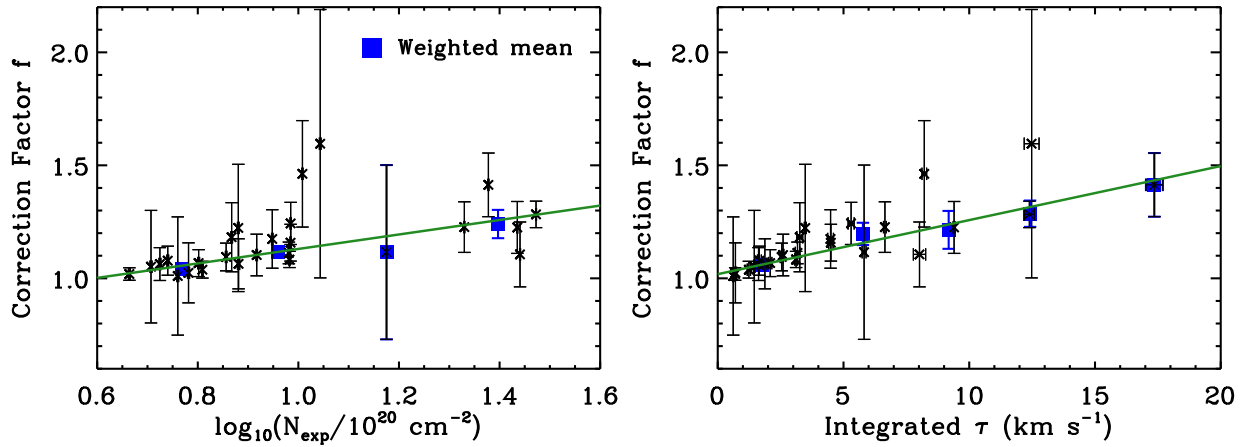}
\caption{\label{f:fcarl} METHOD 1: (left) $f$ = $N_{\rm tot}/N_{\rm exp}$  
($N_{\rm tot}$: derived using the Gaussian decomposition results) as a function of log$_{10}(N_{\rm exp}/10^{20})$.
The blue squares show the (1/$\sigma^{2}$)-weighted mean values in 0.2-wide bins in log$_{10}(N_{\rm exp}/10^{20})$, 
and the linear fit determined for all 26 data points is indicated as the green solid line (Equation \ref{eq:f-carl}). 
(right) $f$ as a function of the integrated optical depth. 
The (1/$\sigma^{2}$)-weighted mean values in 3.4 km s$^{-1}$-wide bins in the integrated optical depth are presented as the blue squares, 
and the green solid line shows the linear fit to all 26 data points:
$f = (0.024 \pm 0.004)\int \tau(v)dv + (1.019 \pm 0.019)$.}
\end{figure*}

\subsection{Calculating the HI Column Density in the Optically Thin Approximation ($N_{\rm low-\tau}$)}
\label{s:low-tau}

As the HI emission spectrum in the direction of a radio continuum source is affected by absorption, 
it is not possible to obtain an emission profile along exactly the same line of sight
as probed by the HI absorption observation. 
For this reason, we instead estimated the ``expected'' HI emission spectrum $T_{\rm exp}(v)$, 
which is the profile we would observe if the continuum source turned off, 
by modeling the ``17-point pattern'' measurements. 
In this modeling, spatial derivatives of the HI emission (up to the second order)      
were carefully taken into account (Section 2.1 of Paper I for details). 
Then the HI column density in the optically thin approximation, $N_{\rm exp}$, can be calculated by 
\begin{equation}
\label{eq:N_exp} 
N_{\textrm{exp}}~(\textrm{cm}^{-2}) = 1.823 \times 10^{18} \int T_{\textrm{exp}}(v)dv~(\textrm{K km s}^{-1}).
\end{equation}
\noindent 
We compute $N_{\rm exp}$ over the velocity range 
where $T_{\rm exp}(v)$ is above its 3$\sigma$ noise level.  
This velocity range covers all CNM and WNM components for each source,  
and the median velocity range for all 26 sources is 
$v_{\rm LSR}$ = $-$39 km s$^{-1}$ to $+$20 km s$^{-1}$.
We then estimate the uncertainty in $N_{\rm exp}$ by propagating 
the $T_{\rm exp}(v)$ error spectrum through Equation (\ref{eq:N_exp}), 
finding a median of $\sim$3.0 $\times$ 10$^{18}$ cm$^{-2}$. 

\begin{figure*}
\centering
\includegraphics[scale=0.5]{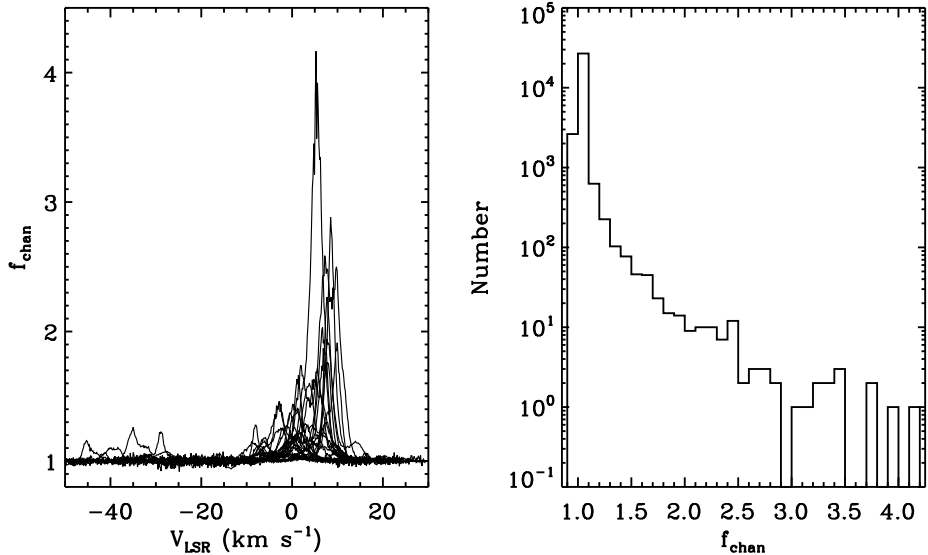}
\caption{\label{f:t_jd} METHOD 2: (left) Correction factor per velocity channel $f_{\rm chan}(v)$ 
(estimated in the isothermal approximation) for all 26 sources. 
(right) Histogram of $f_{\rm chan}(v)$ values.}
\end{figure*}

Additionally, we use spectra from the GALFA-HI cube (pixel size = 1$'$) in the direction of our radio continuum sources 
to derive the ``expected'' emission profiles. 
This simpler approach has been employed 
when HI absorption spectra were obtained without any special strategy such as the ``17-point pattern'' 
(e.g., \citeauthor{Dickey00} 2000; \citeauthor{McClure-Griffiths01} 2001; \citeauthor{Dickey03} 2003). 
We extract HI emission spectra from a 9 $\times$ 9 pixel region  
(roughly the ``17-point pattern'' grid size; Figure 1 of HT03a) centered on each continuum source. 
As the HI spectra right around the continuum source are likely affected by absorption of the background emission, 
we exclude the HI spectra from the central 3 $\times$ 3 pixel region (roughly one Arecibo beamwidth across). 
By averaging the remaining 72 spectra, we then compute an average emission spectrum $T_{\rm avg}(v)$,
and estimate the corresponding HI column density $N_{\rm cube}$ by 
\begin{equation}
\label{eq:N_cube} 
N_{\textrm{cube}}~(\textrm{cm}^{-2}) = 1.823 \times 10^{18} \int T_{\textrm{avg}}(v)dv~(\textrm{K km s}^{-1}).
\end{equation}
\noindent 
The uncertainty in $N_{\rm cube}$ is estimated 
by calculating the standard deviation of the extracted 72 spectra, $\sigma_{T\rm avg}(v)$,  
and propagating it through Equation (\ref{eq:N_cube}).
The median value for all 26 sources is $\sim$1.4 $\times$ 10$^{19}$ cm$^{-2}$. 

In Figure \ref{f:Ncube_Nexp}, we compare the HI column densities calculated using the two methods. 
Most sources probe the HI column density of $\sim$(5--16) $\times$ 10$^{20}$ cm$^{-2}$,
and the last five sources (3C132, 3C131, 3C133, 4C+27.14, and 4C+33.10) extend this range by a factor of $\sim$2.
We find an excellent agreement between the two methods up to $\sim$3 $\times$ 10$^{21}$ cm$^{-2}$. 
The ratio of $N_{\rm exp}$ to $N_{\rm cube}$ ranges from $\sim$0.9 to $\sim$1.1 with a median of $\sim$1.0, 
suggesting that the two HI column density estimates are consistent within 10\%. 
Considering the more careful examination of spatial variations in the HI emission, 
we continue by using $N_{\rm exp}$ as $N_{\rm low-\tau}$ in the following sections.

\subsection{METHOD 1 -- Gaussian Decomposition to Estimate $N_{\rm tot}$}
\label{s:method1} 

In Paper I, we performed Gaussian decomposition of the optical depth and ``expected'' emission spectra, 
and calculated the properties of individual CNM and WNM components  
while considering self-absorption of both the CNM and the WNM by the CNM.  
All Gaussian decomposition results 
(peak brightness temperature, peak optical depth, spin temperature, etc.) 
for each component are presented in Table 2 of Paper I. 
These results enable us to derive the true total HI column density along a line of sight by
\begin{equation}
\label{eq:N-true}
\begin{split}
N_{\textrm{tot}}~(\textrm{cm}^{-2}) & = N_{\textrm{CNM}} + N_{\textrm{WNM}} \\ 
                                    & = 1.823 \times 10^{18} \int (\sum_{0}^{N-1} T_{\textrm{s},n} \tau_{0,n} e^{-[(v-v_{0,n}) / \delta v_{n}]^2} \\  
                                    & \quad + \sum_{0}^{K-1} T_{0,k} e^{-[(v-v_{0,k})/\delta v_{k}]^2})dv~(\textrm{K km s}^{-1}), 
\end{split}  
\end{equation}  
\noindent 
where the components with subscript $n$ refer to the CNM,
the components with subscript $k$ refer to the WNM, 
$\tau_{0}$ is the peak optical depth, 
$v_{0}$ is the central velocity, 
$T_{0}$ is the peak brightness temperature, 
and $\delta v$ is the $1/e$ width of the component.
Here $N_{\rm tot}$ is calculated over the velocity range 
determined as having $T_{\rm exp}(v)$ higher than its 3$\sigma$ noise. 
For the uncertainty in $N_{\rm tot}$, 
we use errors of the fitted parameters provided by Gaussian decomposition to perform a Monte Carlo simulation
where 1000 $N_{\rm CNM}$ and $N_{\rm WNM}$ values are computed from normally distributed parameters. 
The standard deviations of the $N_{\rm CNM}$ and $N_{\rm WNM}$ distributions are then added in quadrature 
to estimate the uncertainty in $N_{\rm tot}$. 
This method of deriving $N_{\rm tot}$ was used by HT03b for their HI absorption measurements 
toward background sources randomly located over the whole Arecibo sky. 
In this study, we focus on a localized group of background sources in the direction of Perseus.

The (integrated) correction factor, $f = N_{\rm tot}/N_{\rm low-\tau} = N_{\rm tot}/N_{\rm exp}$, 
is shown in Figure \ref{f:fcarl} (left) as a function of $N_{\rm exp}$. 
Clearly, the correction factor increases with $N_{\rm exp}$.
We then present the (1/$\sigma^{2}$)-weighted mean values as the blue squares  
and the linear fit to all 26 data points as the green soild line\footnote{In attempting to be consistent 
with our calculation of the uncertainty in $f$ for the isothermal method (Section \ref{s:method2}), 
we run a full Monte Carlo simulation based on errors of the fitted parameters from Gaussian decomposition. 
In this simulation, 1000 $N_{\rm exp}$ and $N_{\rm tot}$ values are calculated from normally distributed parameters, 
and the standard deviation of 1000 $f$ values is used as the uncertainty in $f$. 
We find that linear fit results from using this error estimate,  
$f$ = log$_{10} (N_{\textrm{exp}}/10^{20}) (0.25 \pm 0.03) + (0.87 \pm 0.02)$, 
are consistent with Equation (\ref{eq:f-carl}) within uncertainties.}: 
\begin{equation}
\label{eq:f-carl}
\begin{split}
f & = \log_{10} (N_{\textrm{exp}}/10^{20}) \times a + b \\ 
  & = \log_{10} (N_{\textrm{exp}}/10^{20}) (0.32 \pm 0.06) + (0.81 \pm 0.05).
\end{split}
\end{equation}
\noindent 
In general, the correction factor ranges from $\sim$1.0 at $\sim$3.9 $\times$ 10$^{20}$ cm$^{-2}$ 
to $\sim$1.2 at $\sim$1.3 $\times$ 10$^{21}$ cm$^{-2}$ (maximum uncorrected HI column density in Perseus).  
While $f$ and $N_{\rm exp}$ show a good correlation 
(Spearman's rank correlation coefficient of 0.80),
there are two sources with relatively high correction factors  
at $\sim$10$^{21}$ cm$^{-2}$, 3C092 and 3C093.1. 
Interestingly, the two are located behind the main body of Perseus (Figure 1).
Their high $f$ values of $\sim$1.5--1.6 could result from
an increased amount of the cold HI in the molecular cloud
relative to the surrounding diffuse ISM. 
The CNM fraction is indeed $\sim$0.4 for both sources, 
which is higher than the median value of $\sim$0.3 for all 26 sources (Section 4.3 of Paper I). 
However, this is not the maximum CNM fraction in our measurements ($\sim$0.6). 
Observing a denser grid of radio continuum sources behind Perseus 
and repeating the calculations would be an interesting way to test the cold HI hypothesis.
Finally, the correction factor is also presented as a function of the integrated optical depth in Figure \ref{f:fcarl} (right). 
As expected, there is a clear correlation (Spearman's rank correlation coefficient of 0.94).

We note that our results are not sensitive to HI components at $v_{\rm LSR}$ $<$ $-$20 km s$^{-1}$,      
which are likely unassociated with Perseus (Section 4.2 of Paper I): 
limiting the calculation of $N_{\rm exp}$ and $N_{\rm tot}$ to $v_{\rm LSR}$ $>$ $-$20 km s$^{-1}$ or 
excluding the five sources showing such HI components at large negative velocities 
(corresponding to the sources with log$_{10}(N_{\rm exp}/10^{20})$ $>$ 1.2 in Figure \ref{f:fcarl}; 
3C132, 3C131, 3C133, 4C+27.14, and 4C+33.10)
results in linear fit coefficients that are consistent with what we present here within uncertainties.\footnote{To be specific, 
limiting the calculation of $N_{\rm exp}$ and $N_{\rm tot}$ to $v_{\rm LSR}$ $>$ $-$20 km s$^{-1}$ 
results in $f$ = log$_{10} (N_{\textrm{exp}}/10^{20}) (0.44 \pm 0.07) + (0.73 \pm 0.06)$ 
and $f = (0.034 \pm 0.005) \int \tau(v)dv + (1.004 \pm 0.020)$.
Similarly, excluding the five sources that show the HI components at $v_{\rm LSR}$ $<$ $-$20 km s$^{-1}$ 
leads to $f$ = log$_{10} (N_{\textrm{exp}}/10^{20}) (0.31 \pm 0.11) + (0.82 \pm 0.09)$
and $f = (0.040 \pm 0.011) \int \tau(v)dv + (0.988 \pm 0.027)$.}


\begin{figure*}
\centering
\includegraphics[scale=0.55]{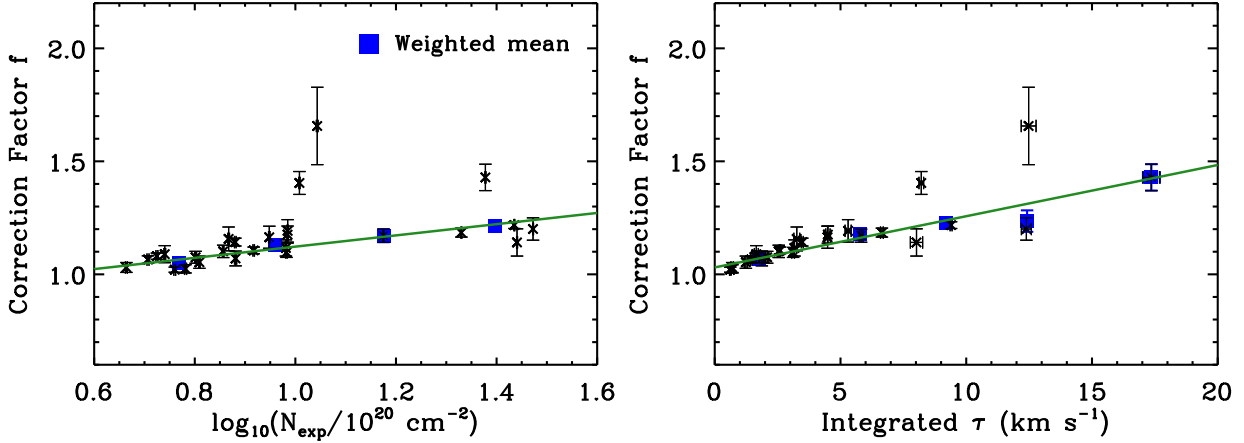}
\caption{\label{f:fjd} METHOD 2: (left) $f$ = $N_{\rm tot}/N_{\rm exp}$ 
($N_{\rm tot}$: estimated in the isothermal approximation) as a function of log$_{10}(N_{\rm exp}/10^{20})$.
The blue squares show the (1/$\sigma^{2}$)-weighted mean values in 0.2-wide bins in log$_{10}(N_{\rm exp}/10^{20})$, 
and the linear fit determined for all 26 data points is indicated as the green solid line (Equation \ref{eq:f-JD}). 
(right) $f$ as a function of the integrated optical depth. 
The (1/$\sigma^{2}$)-weighted mean values in 3.4 km s$^{-1}$-wide bins in the integrated optical depth are presented as the blue squares, 
and the green solid line shows the linear fit to all 26 data points:
$f = (0.023 \pm 0.001)\int \tau(v)dv + (1.030 \pm 0.006)$.}
\end{figure*}

\subsection{METHOD 2 -- Isothermal Estimate of $N_{\rm tot}$}
\label{s:method2}

By assuming that each velocity channel represents gas at a single temperature,
\cite{Dickey00} showed that the correction factor per velocity channel  
can be written as 
\begin{equation}
\label{eq:f-chan1}
f_{\textrm{chan}}(v) = \frac{C_0 T_{\textrm{s}}(v) \tau(v)}{C_0 T_{\textrm{exp}}(v)},
\end{equation}
where $C_0 = 1.823 \times 10^{18}$ cm$^{-2}$/(K km s$^{-1}$). 
In addition, $T_{\rm exp}(v)$ was expressed as 
\begin{equation}
\label{eq:T-exp1}
T_{\textrm{exp}}(v) = T_{\textrm{s}}(v) (1-e^{-\tau(v)}).
\end{equation}
This equation assumes the absence of any radio continuum source behind the absorbing HI cloud. 
As a result, $f_{\rm{chan}}(v)$ simply becomes 
\begin{equation}
\label{eq:f-chan2}
f_{\textrm{chan}}(v) = \frac{\tau(v)}{1-e^{-\tau(v)}}.
\end{equation}

However, while the radio continuum source is absent in Equation (\ref{eq:T-exp1}), 
some diffuse radio continuum emission is always present, and should not be ignored. 
This emission includes the CMB and the Galactic synchrotron emission 
that varies across the sky and becomes strong toward the Galactic plane. 
We call a combination of these contributions as $T_{\rm sky}$, 
and Equation (\ref{eq:T-exp1}) then has to be rewritten as 
\begin{equation}
\label{eq:T-exp2}
T_{\textrm{exp}}^{*}(v) = T_{\textrm{s}}(v) (1-e^{-\tau(v)}) + T_{\textrm{sky}}e^{-\tau(v)}.
\end{equation}
Considering that HI emission spectra are generally baseline subtracted during the reduction process, 
$T_{\rm sky}$ can be removed from both sides of Equation (\ref{eq:T-exp2}). 
Then $T_{\rm exp}(v)$ = $T_{\rm exp}^{*}(v) - T_{\rm sky}$, 
the quantity we have been working with so far, can be expressed as
\begin{equation}
\label{eq:T-exp3}
\begin{split}
T_{\textrm{exp}}(v) & = T_{\textrm{exp}}^{*}(v) - T_{\textrm{sky}} \\
                    & = (T_{\textrm{s}}(v) - T_{\textrm{sky}}) (1-e^{-\tau(v)}).
\end{split}
\end{equation}
As a consequence, the correction factor becomes
\begin{equation}
\label{eq:f-chan3}
f_{\textrm{chan}}(v) = \frac{T_{\textrm{s}}(v)}{T_{\textrm{s}}(v) - T_{\textrm{sky}}} \frac{\tau(v)}{1-e^{-\tau(v)}}, 
\end{equation}
or with direct observables, 
\begin{equation}
\label{eq:f-chan4}
f_{\textrm{chan}}(v) = T_{\textrm{sky}} \frac{\tau(v)}{T_{\textrm{exp}}(v)} + \frac{\tau(v)}{1-e^{-\tau(v)}}.
\end{equation}

\begin{figure*}
\centering 
\includegraphics[scale=0.25]{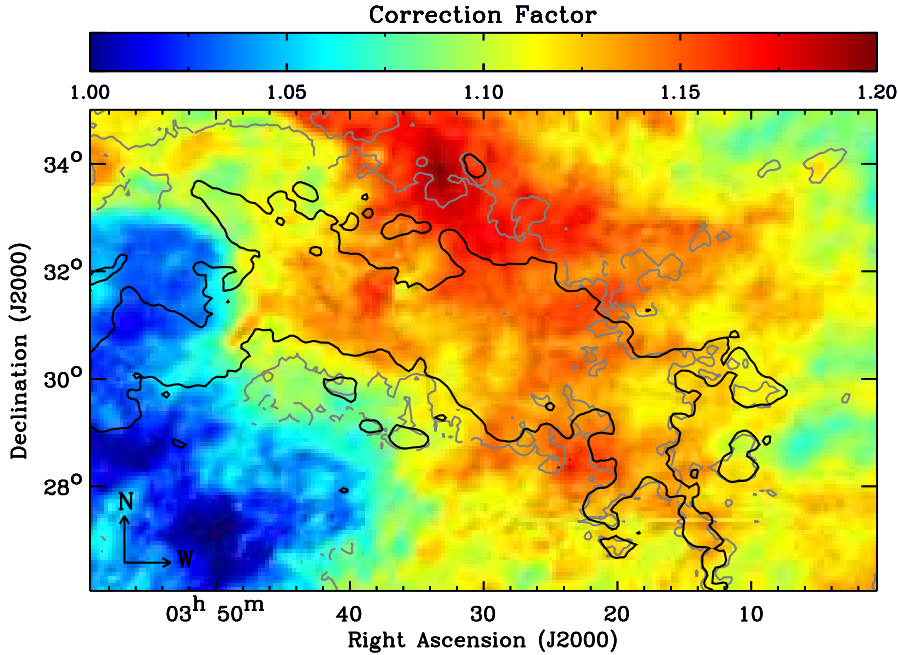}
\caption{\label{f:corr-fact} Correction factor at 4.3$'$ resolution 
estimated in Section \ref{s:method1} (Equation \ref{eq:f-carl}). 
In addition, the 3$\sigma$ contour of the ``old'' $N$(H$_{2}$) from \citet{Lee12} 
(before the correction for high optical depth) is overlaid in gray, 
while the 3$\sigma$ contour of the CfA $I_{\rm CO}$ is shown in black. 
The resolutions of the $N$(H$_{2}$) and $I_{\rm CO}$ images are 4.3$'$ and 8.4$'$.}
\end{figure*}

In order to estimate the contribution from the Galactic synchrotron emission, 
we use the \cite{Haslam82} 408 MHz survey of the Galaxy. 
The brightness temperature at 408 MHz is converted to 1.4 GHz using the spectral index of $-$2.7. 
As the absolute Galactic latitude of our continuum sources is generally higher than 10$^{\circ}$, 
the synchrotron contribution is small with $T_{\rm sky}$ ranging from 2.78 K to 2.80 K (Table 1 of Paper I).  
Based on the histogram of $T_{\rm s}$ for the individual CNM components (Figure 5b of Paper I), 
we can then provide a rough estimate of $T_{\rm s}(v)/[T_{\rm s}(v)-T_{\rm sky}]$:
the expected range is narrow, from $\sim$1.0 to $\sim$1.2.
Clearly, for molecular clouds located closer to the Galactic plane 
the contribution from the diffuse radio continuum emission will be more significant. 

In Equation (\ref{eq:f-chan1}), $f_{\rm chan}(v)$ essentially represents the correction 
that needs to be applied to $T_{\rm exp}(v)$ to calculate the true brightness temperature profile.
As a result, the true total HI column density can be obtained by 
\begin{equation}
\label{eq:N-true-JD}
N_{\textrm{tot}}~(\textrm{cm}^{-2}) = 1.823 \times 10^{18} \int f_{\textrm{chan}}(v) T_{\textrm{exp}}(v)dv~(\textrm{K km s}^{-1}). 
\end{equation}
We derive $f_{\rm chan}(v)$ for all 26 sources (Equation \ref{eq:f-chan4}), 
and present the results in Figure \ref{f:t_jd}.
About 91\% of the $f_{\rm chan}(v)$ values are between 1 and 2, 
and the fraction of velocity channels with $f_{\rm chan}(v) > 2$ is very small.
We then calculate $N_{\rm tot}$ using Equation (\ref{eq:N-true-JD}), 
and show the (integrated) correction factor, $f = N_{\rm tot}/N_{\rm exp}$,
as a function of $N_{\rm exp}$ in Figure \ref{f:fjd} (left).
Here the integration is done over the velocity range 
where $T_{\rm exp}(v)$ is higher than its 3$\sigma$ noise.
The uncertainty in $f$ is derived by running a Monte Carlo simulation 
where the optical depth and ``expected'' emission error spectra 
are propagated through Equations (\ref{eq:f-chan4}) and (\ref{eq:N-true-JD}) to compute 1000 $f$ values. 
The standard deviation of the $f$ distribution is used as the final uncertainty in $f$.  

Similar to the Gaussian decomposition method, 
we find a good correlation between $f$ and $N_{\rm exp}$ (Spearman's rank correlation coefficient of 0.84). 
The linear fit determined using all 26 data points is 
\begin{equation}
\label{eq:f-JD}
f = \log_{10}(N_{\textrm{exp}}/10^{20})(0.25 \pm 0.02) + (0.87 \pm 0.02).
\end{equation}
Additionally, $f$ is plotted as a function of the integrated optical depth in Figure \ref{f:fjd} (right), 
again showing a clear correlation (Spearman's rank correlation coefficient of 0.97). 

Both graphs in Figure \ref{f:fjd} are very similar with 
those in Figure \ref{f:fcarl} for the Gaussian decomposition method. 
Specifically, the linear fit coefficients are consistent within uncertainties.  
This is surprising considering that the two methods are very different. 
In particular, the isothermal method assigns a single spin temperature to each velocity channel, 
while the Gaussian decomposition method allows a single velocity channel to have contributions 
from several HI components with different spin temperatures.

In \cite{Dickey00}, the authors updated the isothermal method 
by incorporating the two-phase approximation. 
As input parameters, this method then required the spin temperature of the cold HI 
and the fraction of the warm HI that is in front of the cold HI, 
the quantity they referred to as $q$. 
\cite{Dickey00} showed that for their SMC data  
there is no difference between the one- and two-phase approximations regarding the correction factor if $q \gtrsim 0.5$, 
while the difference becomes more pronounced when $q \lesssim 0.25$.
In the Gaussian decomposition method, this fraction ($F_k$ in Paper I) 
is important as well, but is difficult to constrain. 
Thus, the fitting process was repeated for $F_k$ = 0, 0.5, and 1, 
and these results were used to estimate the final uncertainty in $T_{\rm s}$ 
(Section 3 of Paper I and HT03a for details).

To understand why the two different methods result in similar correction factors, 
we compare Equations (\ref{eq:N-true}) and (\ref{eq:N-true-JD}) (Appendix A), 
and find that they become comparable regardless of $F_k$ when $\tau \ll 1$ and $T_{\rm sky} \ll T_{\rm s}$. 
In our observations of Perseus, the median peak optical depth for all CNM components is $\sim$0.2,  
and only a small number of the components has the peak optical depth higher than 1 
(10 out of 107; Section 4.1 of Paper I), satisfying the condition.
In addition, we already showed that $T_{\rm sky}$ is small with $\sim$2.8 K for Perseus.
The difference between the Gaussian decomposition and isothermal methods, however, will be more significant 
for molecular clouds that have a large amount of the cold, optically thick HI and/or 
a substantial contribution from the diffuse radio continuum emission.  
Due to the more self-consistent way to derive $T_{\rm exp}(v)$, 
we continue by using the Gaussian decomposition results for further analyses.

Finally, we note that Equations (\ref{eq:f-carl}) and (\ref{eq:f-JD}) 
could be biased against very high optical depths as they result in saturated absorption spectra,
e.g., 4C$+$32.14, the source we had to exclude from our analyses due to its highly saturated absorption profile.
In addition, the equations are based on our explicit assumption of the linear relation between $f$ and log$_{10}$($N_{\rm exp}/10^{20}$). 
The HI emission and absorption measurements obtained by HT03a,b along many random lines of sight through the Galaxy 
show that the linear relation indeed describes the observations well up to log$_{10}$($N_{\rm exp}/10^{20}$) $\sim$ 1.5, 
and the fitted coefficients are consistent with Equations (\ref{eq:f-carl}) and (\ref{eq:f-JD}) within uncertainties (Appendix B).
There is some interesting deviation from the linear relation, however, 
particularly for a few sources with log$_{10}$($N_{\rm exp}$/10$^{20}$) $\gtrsim$ 1. 
This deviation could suggest a non-linear relation at high column densities, 
and its significance needs to be further examined with more HI absorption measurements. 
In the future, it will also be important to study the dense CNM using alternative tracers 
such as CI and CII fine-structure lines (e.g., \citeauthor{Pineda13} 2013).

\begin{figure*} 
\centering
\includegraphics[scale=0.25]{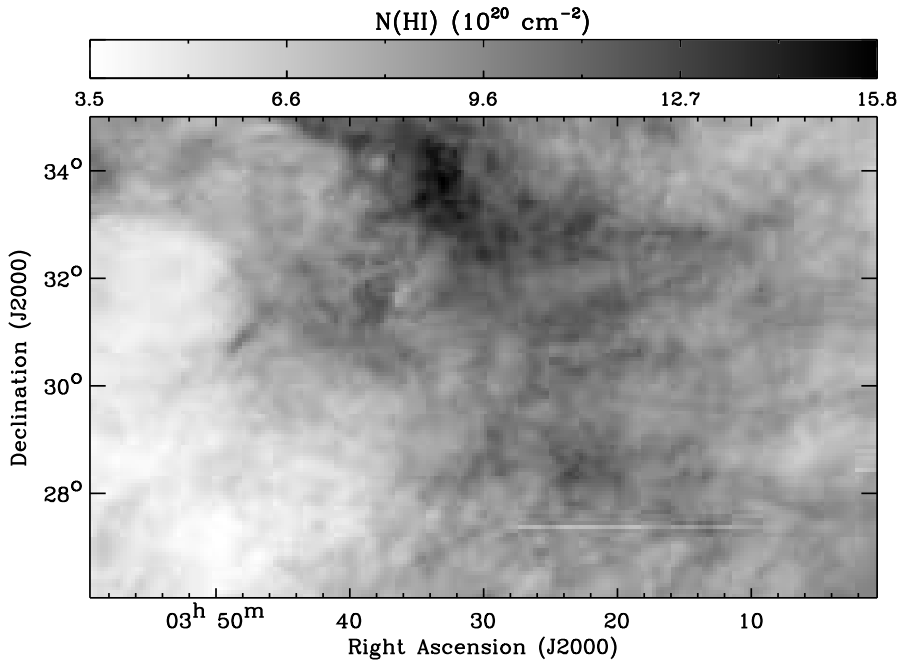}
\caption{\label{f:corrected-HI} Corrected $N$(HI) image at 4.3$'$ resolution.}
\end{figure*}



\subsection{Comparison with Previous Studies}
\label{s:comparison}

\textit{Dickey et al. (2000)}. 
The correction factor calculated by \cite{Dickey00} for the SMC 
using full line of sight information can be rewritten as 
$f = \log_{10} (N_{\rm exp}/10^{20}) 0.667 + 0.066$ for $N_{\rm exp} > 2.5 \times 10^{21}$ cm$^{-2}$. 
The range of the HI column density we probe for Perseus barely overlaps with that in \cite{Dickey00}, 
as the HI column density in the low-metallicity SMC is significantly higher compared to the Galaxy. 
While the difference with \cite{Dickey00} in $f$ values depends on $N_{\rm exp}$, 
our correction factor is only $\sim$4\% higher than what \cite{Dickey00} suggests 
when extrapolated to their maximum HI column density of 10$^{22}$ cm$^{-2}$. 
Similarly, \cite{Dickey03} used HI emission and absorption data 
from the Southern Galactic Plane Survey (\citeauthor{McClure-Griffiths05} 2005)
in combination with the isothermal method, 
and estimated $f$ = $\sim$1.4--1.6 for sources 
located at 326$^{\circ}$ $< l <$ 333$^{\circ}$ and $|b|$ $\lesssim$ 1$^{\circ}$. 
This correction factor is comparable with what we find for Perseus.

\textit{Heiles \& Troland (2003a,b)}. HT03a,b performed Gaussian decomposition 
of 79 HI emission/absorption spectral line pairs, 
and derived the correction factor, which they called $R_{\rm raw} = 1/f$. 
They found that $R_{\rm raw}$ ranges from $\sim$0.3 to $\sim$1.0, 
corresponding to $f$ = $\sim$1.0--3.0 (Appendix B). 
In particular, they estimated $f \sim 1.3$ for the Taurus/Perseus region,
which is similar with what we find for Perseus. 

\textit{Braun et al. (2009)}. Our correction factor is 
smaller than what \cite{Braun09} and \cite{Braun12} 
claimed for M31, M33, and the LMC based on the modeling of HI emission spectra. 
They found that the correction exceeds an order of magnitude in many cases, 
and increases the global HI mass by $\sim$30\%. 
Even when considering the correction factor per velocity channel, 
we find the maximum $f_{\rm chan}(v)$ $\sim$ 4 for only one source 
and $f_{\rm chan}(v) \lesssim 3$ for the rest of our sources.

\textit{Chengalur et al. (2013)}. Our correction factor 
can also be compared with predictions from \cite{Chengalur13}. 
This study performed Monte Carlo simulations 
where observationally motivated input parameters such as the column density, 
the spin temperatures of the CNM and the WNM, and the fraction of gas in each of the different phases 
were provided for ISM models.
Our correction factor versus integrated optical depth plots, Figures \ref{f:fcarl} (right) and \ref{f:fjd} (right), 
can be directly compared with Figure 1A in \cite{Chengalur13}. 
We find that the correction factor by \cite{Chengalur13} is significantly higher than our estimate, 
although the general trend of increasing correction with the integrated optical depth is similar. 
For example, \cite{Chengalur13} expects $f \sim$ 20 when $\int \tau dv$ $\sim$ 10 km s$^{-1}$, 
while we find only $f < 1.5$. 
Similarly, our Figures \ref{f:fcarl} (left) and \ref{f:fjd} (left) 
can be compared with Figure 2A in \cite{Chengalur13}. 
We find that our estimate is consistent with the correction factor by \cite{Chengalur13} 
for the column density less than 10$^{21}$ cm$^{-2}$, 
while \cite{Chengalur13} overestimates at the high end of our column density range. 
If we extrapolate our relation to 10$^{22}$ cm$^{-2}$, 
we expect a $\sim$10 times lower correction factor than what \cite{Chengalur13} suggests.
The reason for their very high correction factor could be the 
inclusion of extremely high column densities (10$^{23}$--10$^{24}$ cm$^{-2}$) in their ISM models, 
although it is not clear why it less affects the isothermal estimate of the HI column density. 
In the simulations, the median ratio of the true HI column density to the isothermal estimate was $\sim$1.1 
when $\int \tau dv$ $\sim$ 5 km s$^{-1}$.   

\begin{figure}
\centering 
\includegraphics[scale=0.42]{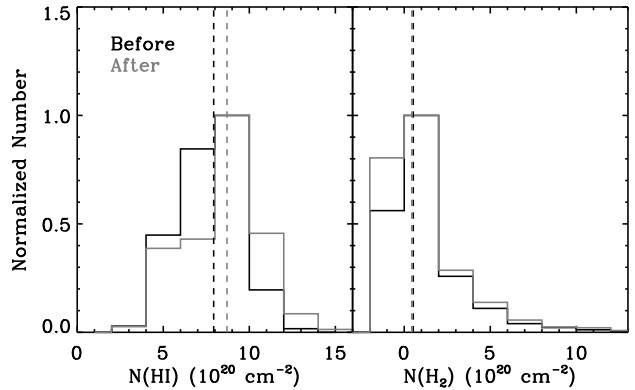}
\caption{\label{f:HI-H2-comp} (left) Normalized histograms of the two $N$(HI) images, 
before (black) and after (gray) the correction for high optical depth. 
The median of each histogram is shown as the dashed line. 
(right) Same as the left panel, but for $N$(H$_{2}$).}
\end{figure}

\begin{figure*} 
\centering 
\includegraphics[scale=0.25]{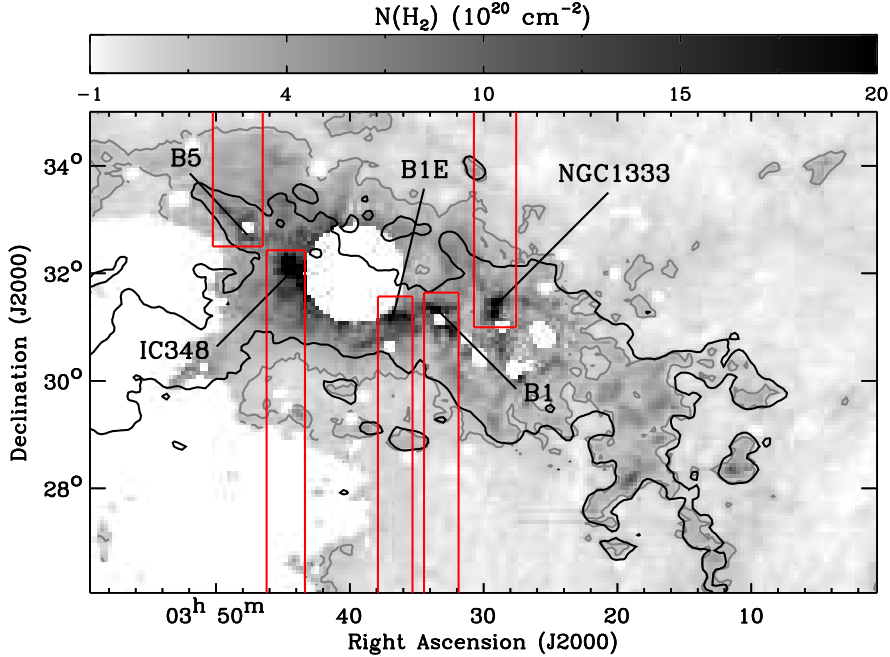}
\caption{\label{f:new-H2} Rederived $N$(H$_{2}$) image.
The blank pixels correspond to regions with possible contaminations 
(point sources, the Taurus molecular cloud, and the ``warm dust ring''; Section 4 of \citeauthor{Lee12} 2012 for details). 
The 3$\sigma$ contour of the new $N$(H$_{2}$) is overlaid in gray, 
while the 3$\sigma$ contour of the CfA $I_{\rm CO}$ is shown in black.
The resolutions of the $N$(H$_{2}$) and $I_{\rm CO}$ images are 4.3$'$ and 8.4$'$. 
The red rectangular boxes show the boundaries of the selected dark (B5, B1E, and B1) and star-forming (IC348 and NGC1333) regions.}
\end{figure*}

\textit{Liszt (2014)}. Using the HI absorption data compiled by \cite{Liszt10}, 
\cite{Liszt14b} estimated the correction factor for radio continuum sources located at high Galactic latitudes. 
While they did not provide detailed information about how exactly the derivation was done,  
their correction factor was small with $f$ less than 1.2 for $E(B-V)$ $\lesssim$ 0.5 mag.  
This is comparable to our finding. 

\textit{Fukui et al. (2014,2015)}. Finally, \cite{Fukui15} 
estimated the correction factor for the Galaxy at $|b|$ $>$ 15$^{\circ}$   
by exploring the relation between $\tau_{353}$ and $N$(HI) at 33$'$ resolution.  
Their Figure 13 shows that the correction factor distribution 
ranges from $\sim$1.0 to $\sim$3.0, and peaks at $\sim$1.5. 
Using the same methodology, \cite{Fukui14} found that the correction increases the total HI mass of 
the high latitude molecular clouds MBM 53, 54, 55, and HLCG 92-35 by a factor of $\sim$2. 
In general, the correction factor by Fukui et al. (2014,2015) appears higher than our estimate for Perseus.   
While we do not perform a detailed comparison with Fukui et al. (2014,2015),  
we test their claim of the optically thick HI as an alternative of the ``CO-dark'' gas in Section \ref{s:CO-dark}.   

\section{Applying the Correction for High Optical Depth to Perseus} 
\label{s:perseus-corrected} 
We apply the correction derived in Section \ref{s:method1} 
to the $N$(HI) image of Perseus from \cite{Lee12} 
for pixels with log$_{10}$($N_{\rm exp}$/10$^{20}$) $>$ 0.6 
(where the correction factor is higher than 1). 
The correction factor and corrected $N$(HI) images are shown  
in Figures \ref{f:corr-fact} and \ref{f:corrected-HI}. 
In addition, we present the normalized histograms of the two $N$(HI) images, 
before and after the correction, in Figure \ref{f:HI-H2-comp}. 

As Figure \ref{f:HI-H2-comp} indicates, 
the correction does not make a significant change in $N$(HI). 
To be specific, the median $N$(HI) increases by a factor of $\sim$1.1 
from $\sim$7.9 $\times$ 10$^{20}$ cm$^{-2}$ to $\sim$8.7 $\times$ 10$^{20}$ cm$^{-2}$, 
while the maximum $N$(HI) increases by a factor of $\sim$1.2 
from $\sim$1.3 $\times$ 10$^{21}$ cm$^{-2}$ to $\sim$1.6 $\times$ 10$^{21}$ cm$^{-2}$. 
In terms of the total HI mass, the correction results in a $\sim$10\% increase 
from $\sim$2.3 $\times$ 10$^{4}$ M$_{\odot}$ to $\sim$2.5 $\times$ 10$^{4}$ M$_{\odot}$. 
This increase in the HI mass is comparable to what \citet{Dickey00} found for the SMC, 
but is smaller than the value estimated by \citet{Braun09} and \citet{Braun12} for M31, M33, and the LMC.
In addition, we note that our correction factor image looks smooth 
compared to a granulated appearance of the corrected $N$(HI) images by \citet{Braun09} and \citet{Braun12}, 
although Perseus ($\sim$80 pc $\times$ 50 pc)
would be unresolved or only marginally resolved in their studies.   
Finally, the HI mass increase due to the optically thick gas in Perseus is  
smaller than what Fukui et al. (2014,2015) derived for Galactic molecular clouds. 

\section{How Does the High Optical Depth Affect the HI Saturation in Perseus?}
\label{s:revisit-saturation}

\subsection{Rederiving $N$\textup{(H$_{2}$)}}
\label{s:H2} 
To investigate the impact of high optical depth on the observed HI saturation in Perseus, 
we first rederive the $N$(H$_{2}$) image using the corrected $N$(HI).  
In essence, we use the same methodology as \citet{Lee12}: 
the $A_{V}$ image is derived using the IRIS 60/100 $\mu$m and 2MASS $A_{V}$ data, 
and a local D/G is adopted to estimate $N$(H$_{2}$). 
We refer to Section 4 of \citet{Lee12} for details on the method for deriving $N$(H$_{2}$) and its limitation, 
and summarize here main results. 

\begin{figure*}
\centering
\includegraphics[scale=0.33]{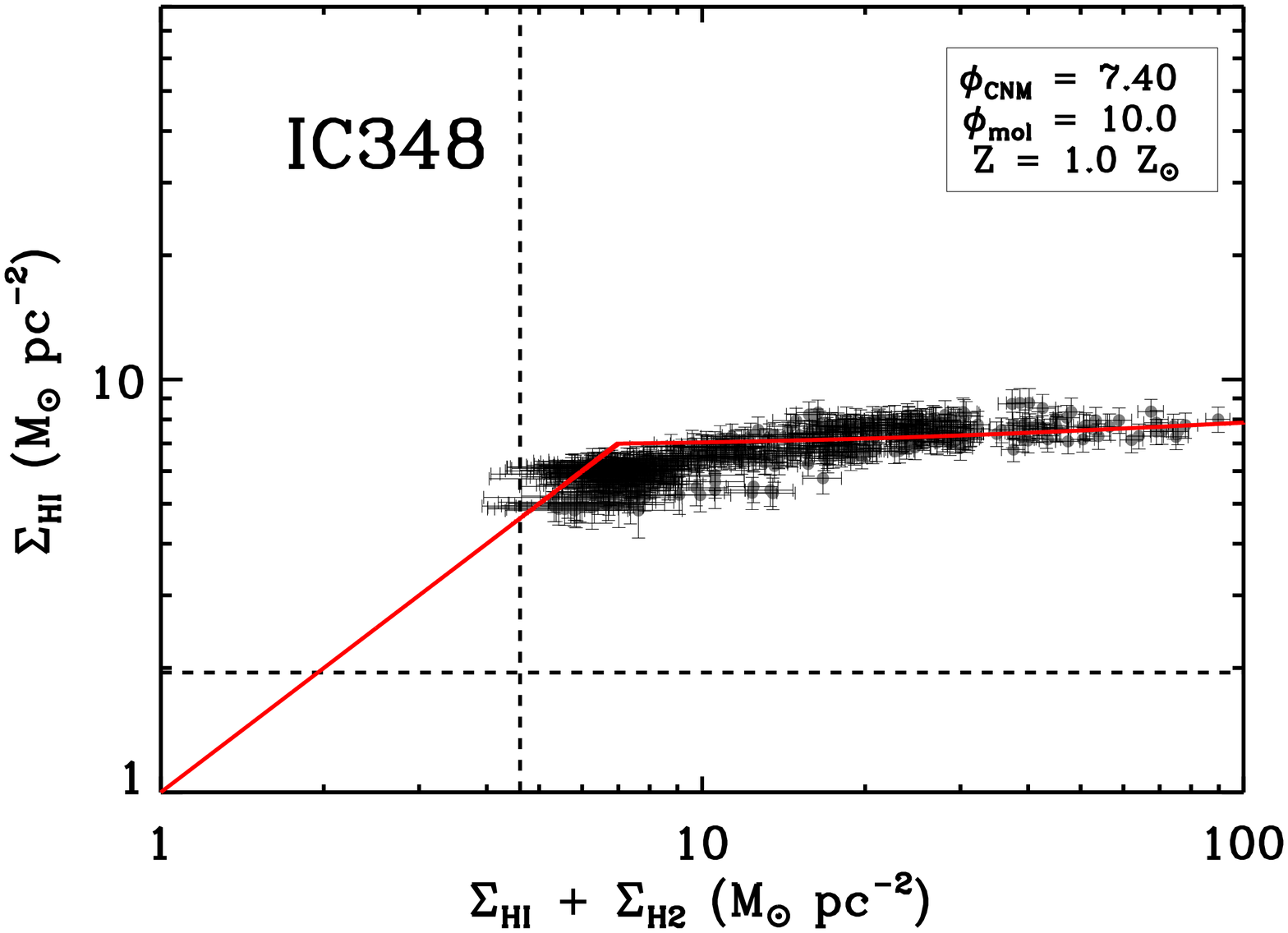}
\includegraphics[scale=0.33]{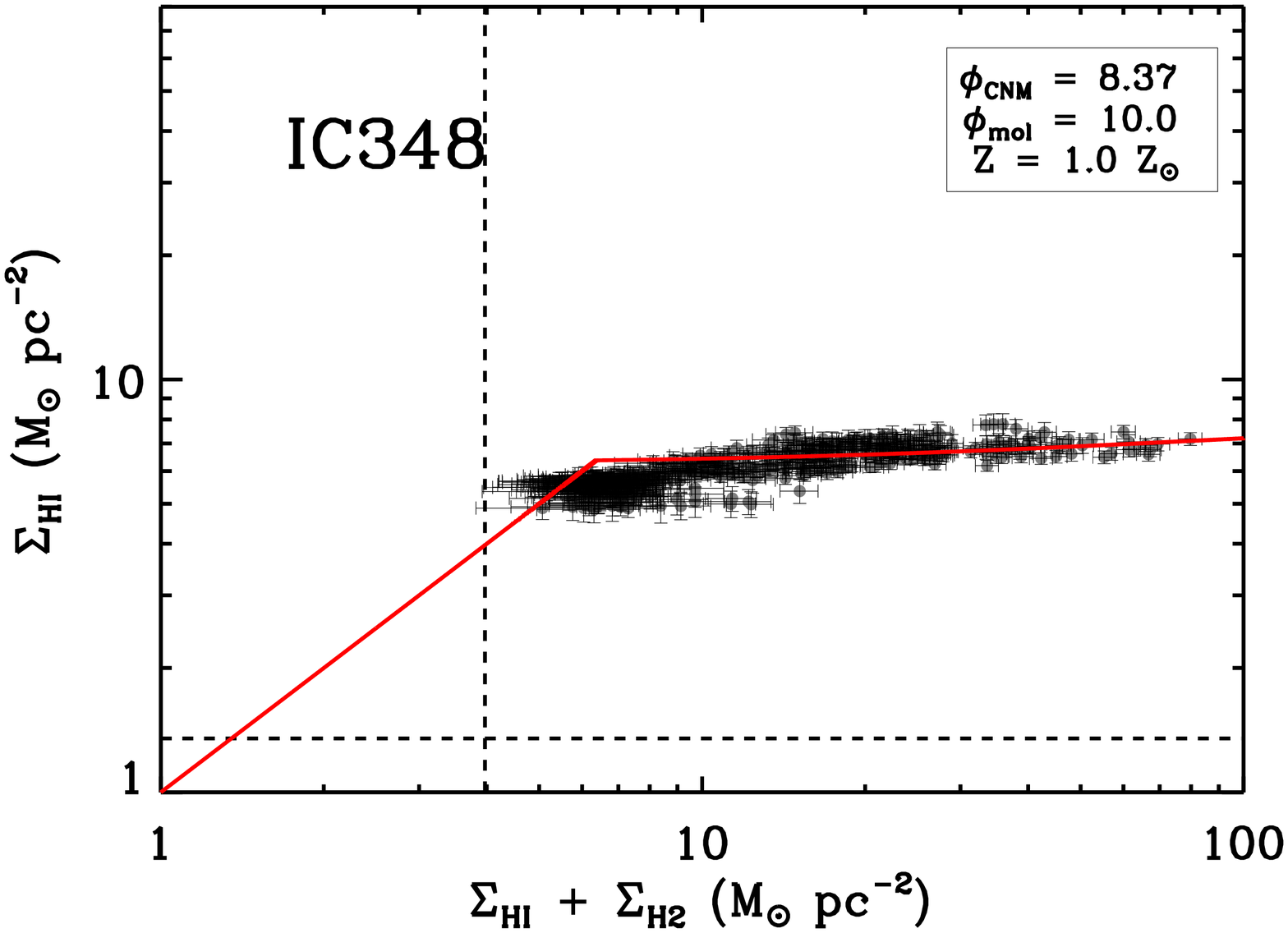}
\includegraphics[scale=0.33]{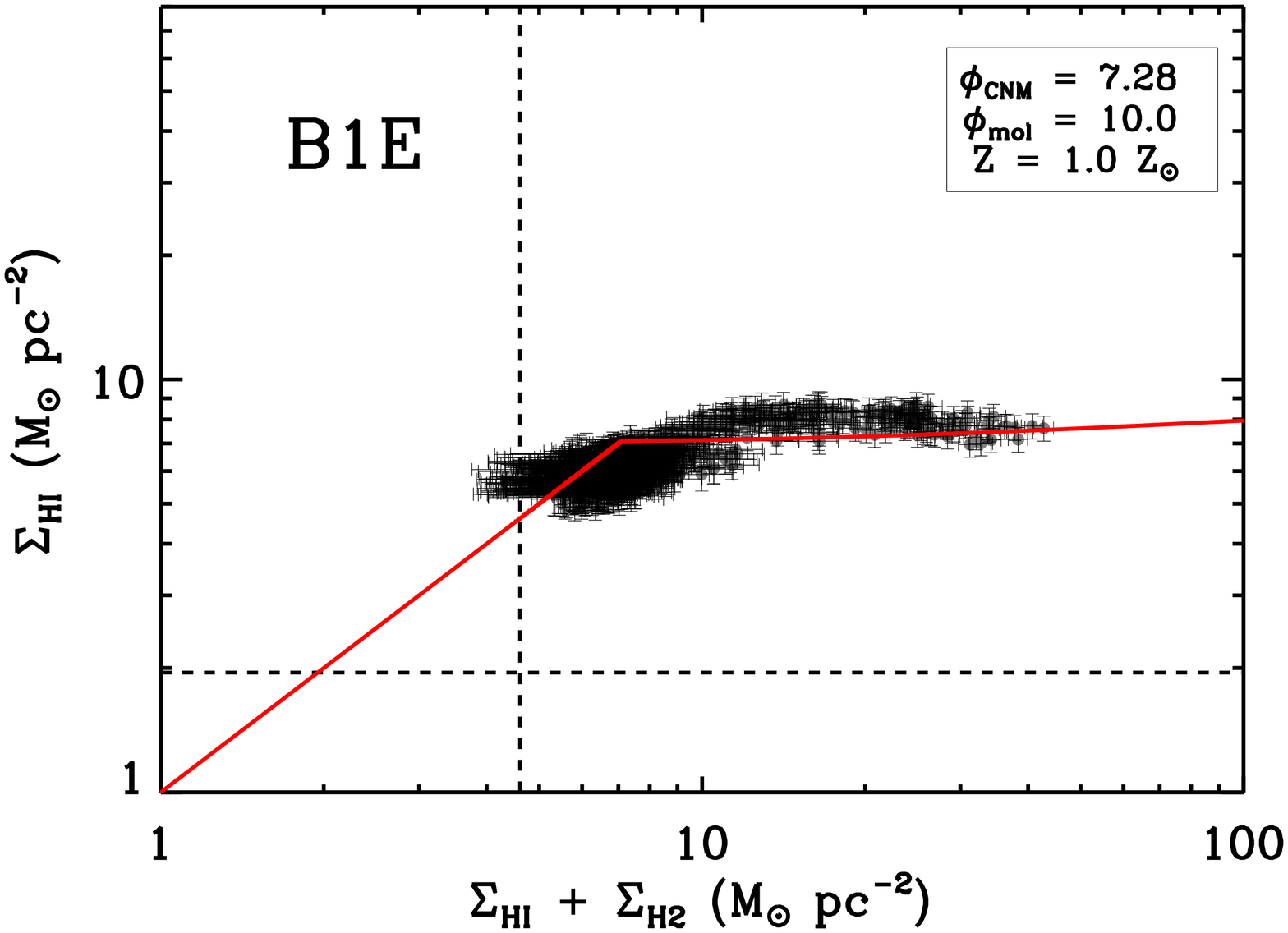}
\includegraphics[scale=0.33]{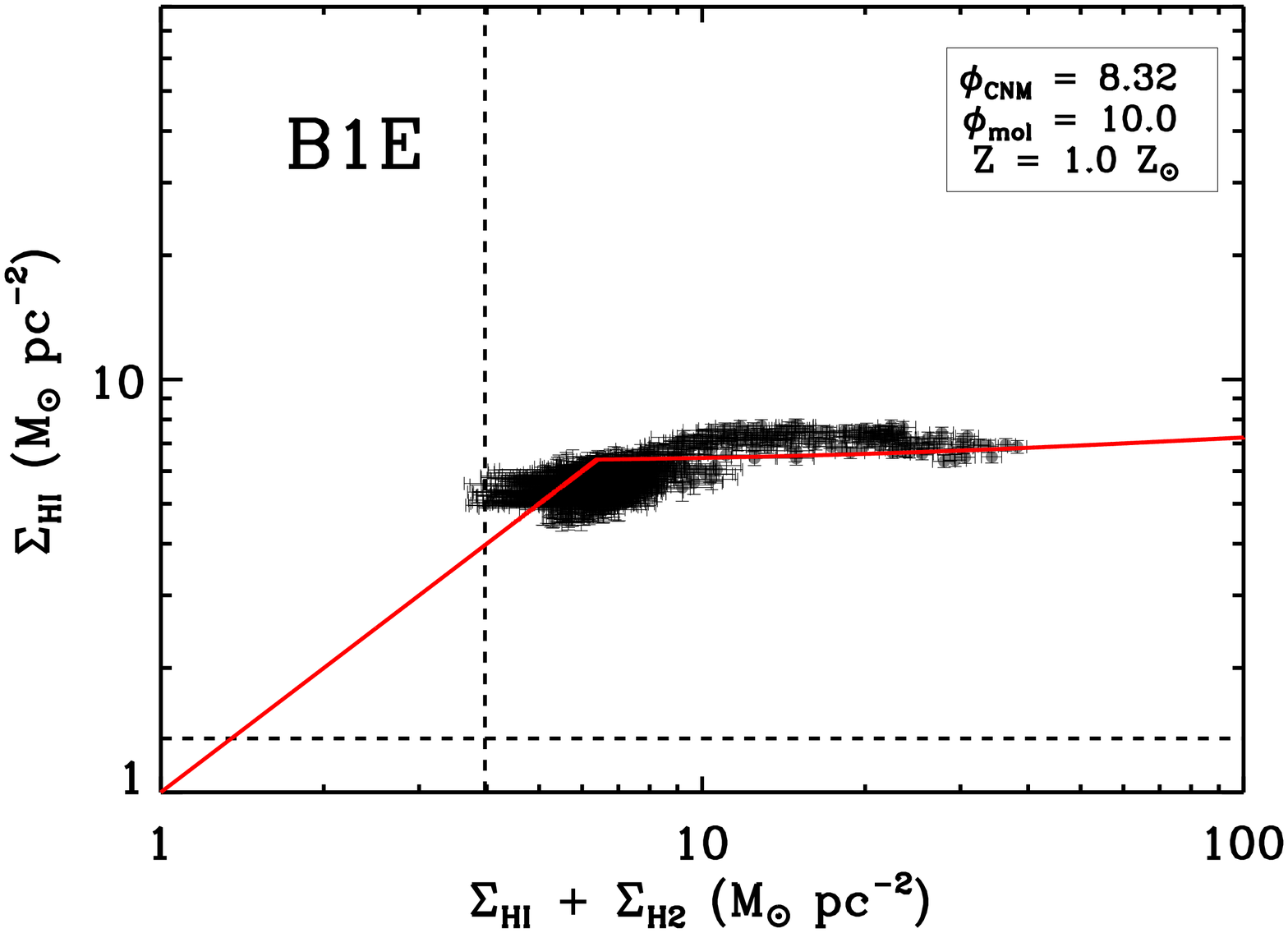}
\caption{\label{f:HI} (left) $\Sigma_{\rm HI}$ versus $\Sigma_{\rm HI}+\Sigma_{\rm H2}$ for IC348 and B1E (this study).  
All finite pixels in the rectangular boxes shown in Figure \ref{f:new-H2} are used for plotting.
The median 3$\sigma$ values of $\Sigma_{\rm HI}$ and $\Sigma_{\rm HI}+\Sigma_{\rm H2}$ are indicated as the black dashed lines, 
while the best-fit curves determined in Section \ref{s:KMT09} are overlaid in red.
The best-fit parameters are shown in the top right corner of each plot.
(right) Plots from \citet{Lee12}.} 
\end{figure*}

1. Contamination from VSGs:  
To estimate $T_{\rm dust}$ from the ratio of $I_{60}$ to $I_{100}$, 
the contribution from stochastically heated VSGs to $I_{60}$ must be removed. 
For this purpose, we compare our IRIS-based $T_{\rm dust}$ with the DIRBE-based $T_{\rm dust}$ from \citet{Schlegel98}, 
and find that the contribution from VSGs to $I_{60}$ is 78\%. 
This is the same as what \citet{Lee12} found. 

2. Zero point calibration for $\tau_{100}$: 
We refine the zero point of the $\tau_{100}$ image by assuming that 
the dust column density traced by $\tau_{100}$ is proportional to $N$(HI) for atomic-dominated regions. 
Based on the zero point of the $\tau_{100}$--$N$(HI) relation, 
we add 1.1 $\times$ 10$^{-4}$ to the $\tau_{100}$ image. 
This is slightly smaller than what \cite{Lee12} added, i.e., 1.8 $\times$ 10$^{-4}$.

3. Conversion from $\tau_{100}$ to $A_{V}$:
We convert $\tau_{100}$ into $A_{V}$ by adopting $X = 740$ for $A_{V}$ = $X \tau_{100}$ 
that results in the best agreement between our IRIS-based $A_{V}$ and the 2MASS-based $A_{V}$ from \citet{Ridge06b}.
This is slightly higher than $X = 720$ used by \citet{Lee12}. 
We compare our rederived $A_{V}$ with the $A_{V}$ image from \citet{Lee12}, 
and find that the ratio of the new $A_{V}$ to the old $A_{V}$ ranges from $\sim$0.93 to $\sim$1.02.

4. Deriving a local D/G and $N$(H$_{2}$): 
We examine the $A_{V}$--$N$(HI) relation, and find that the slope of $A_{V}$/$N$(HI) = 1 $\times$ 10$^{-21}$ mag cm$^{2}$ 
is a good measure of D/G for Perseus. 
This is slightly lower than what \citet{Lee12} estimated, i.e., 1.1 $\times$ 10$^{-21}$ mag cm$^{2}$, 
and makes sense considering that our rederived $A_{V}$ is essentially 
the same with the $A_{V}$ image from \citet{Lee12}, 
while the corrected $N$(HI) is slightly higher than the uncorrected $N$(HI). 
Finally, we derive $N$(H$_{2}$) using Equation (\ref{eq:H2}), 
and mask pixels with possible contaminations 
(point sources, the Taurus molecular cloud, and the ``warm dust ring''), 
following what \citet{Lee12} did. 
We show the final $N$(H$_{2}$) image in Figure \ref{f:new-H2},
as well as the normalized histograms of the new (this study) 
and old \citep{Lee12} $N$(H$_{2}$) images in Figure \ref{f:HI-H2-comp} (right).

Conclusion: In our rederivation of $N$(H$_{2}$), 
all parameters are identical with or only slightly different from what \citet{Lee12} used. 
As a result, the new $N$(H$_{2}$) is comparable to the old $N$(H$_{2}$), 
shown as a good agreement between the two histograms in Figure \ref{f:HI-H2-comp} (right). 
The rederived $N$(H$_{2}$) ranges from 
$-$1.5 $\times$ 10$^{20}$ cm$^{-2}$ to 5.1 $\times$ 10$^{21}$ cm$^{-2}$ 
with a median of 4.6 $\times$ 10$^{19}$ cm$^{-2}$,
and $\sim$83\% of the pixels whose S/N is higher than 1 
have the new $N$(H$_{2}$) differing from the old $N$(H$_{2}$) by only 10\%. 
Finally, we note that $\sim$30\% of all finite pixels 
have negative $N$(H$_{2}$) values of mostly around $-$(1--5) $\times$ 10$^{19}$ cm$^{-2}$, 
which are very close to zero considering 
the median uncertainty in $N$(H$_{2}$) of $\sim$5 $\times$ 10$^{19}$ cm$^{-2}$ (Section \ref{s:errors}).  

\subsection{Uncertainty in $N$\textup{(H$_{2}$)}}
\label{s:errors}
As \citet{Lee12} did, we perform a series of Monte Carlo simulations 
to estimate the uncertainty in $N$(H$_{2}$). 
In these simulations, we assess the errors in $N$(HI) and $A_{V}$, and 
propagate them together. 

\begin{figure*} 
\centering
\includegraphics[scale=0.33]{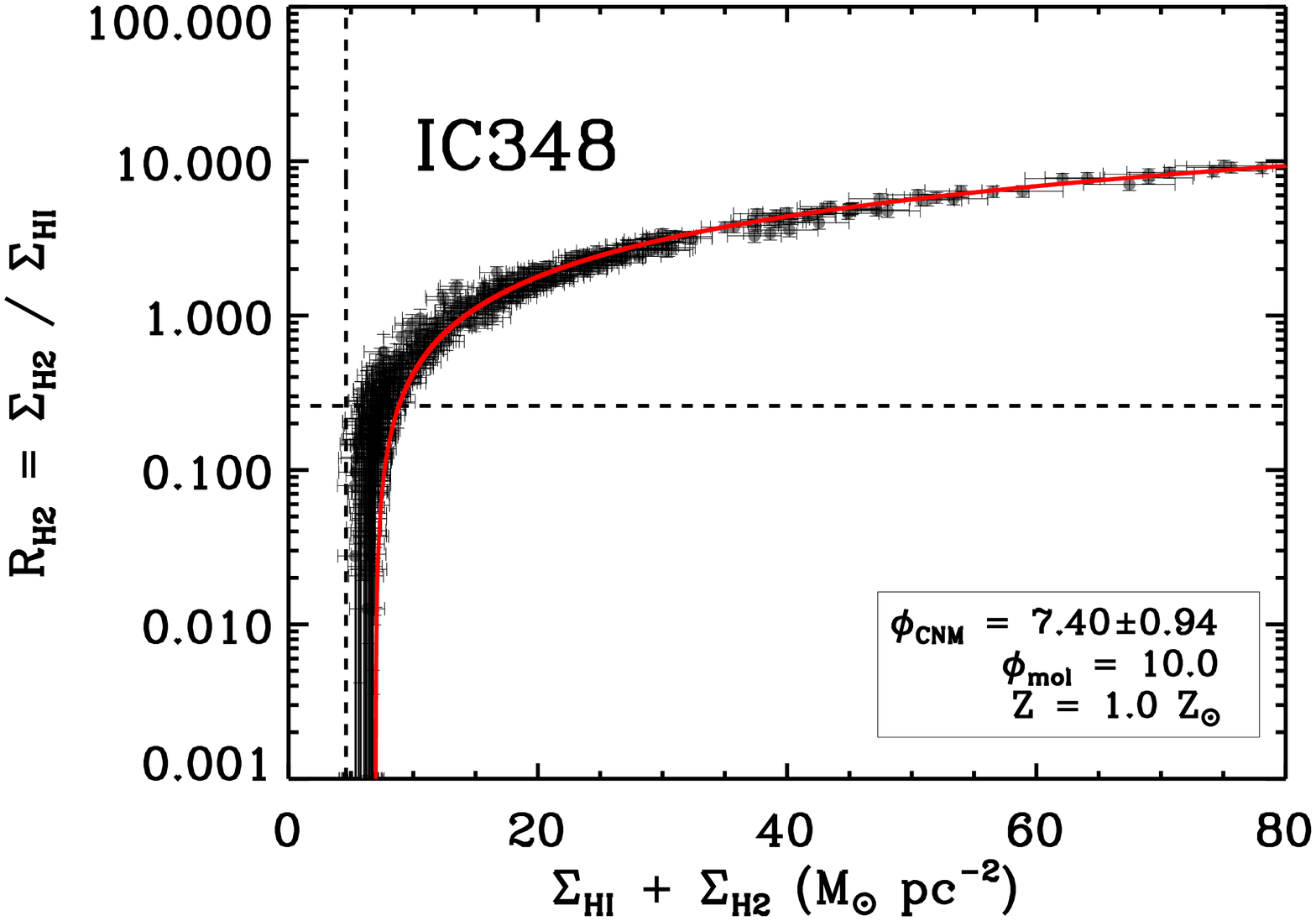}
\includegraphics[scale=0.33]{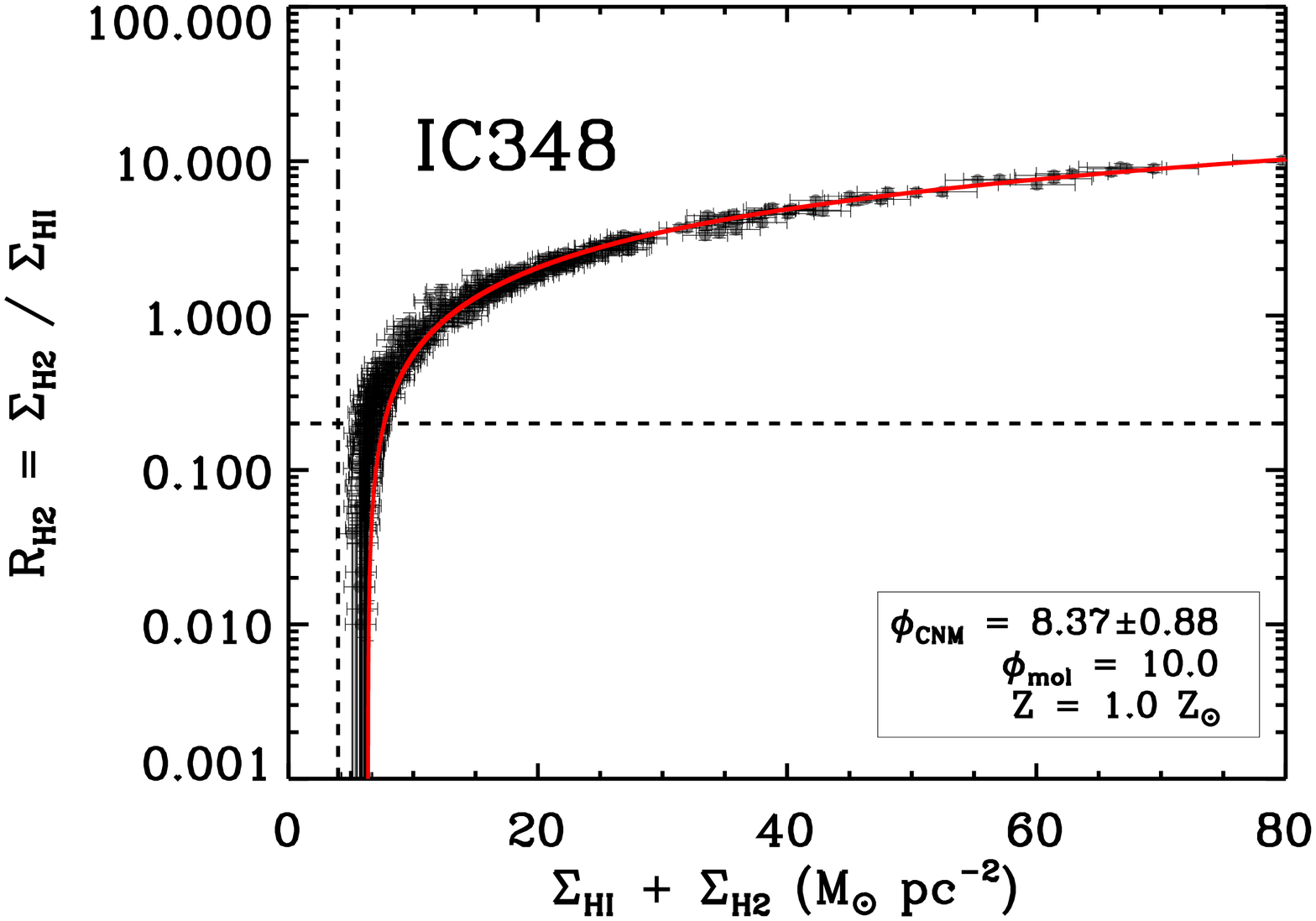}
\includegraphics[scale=0.33]{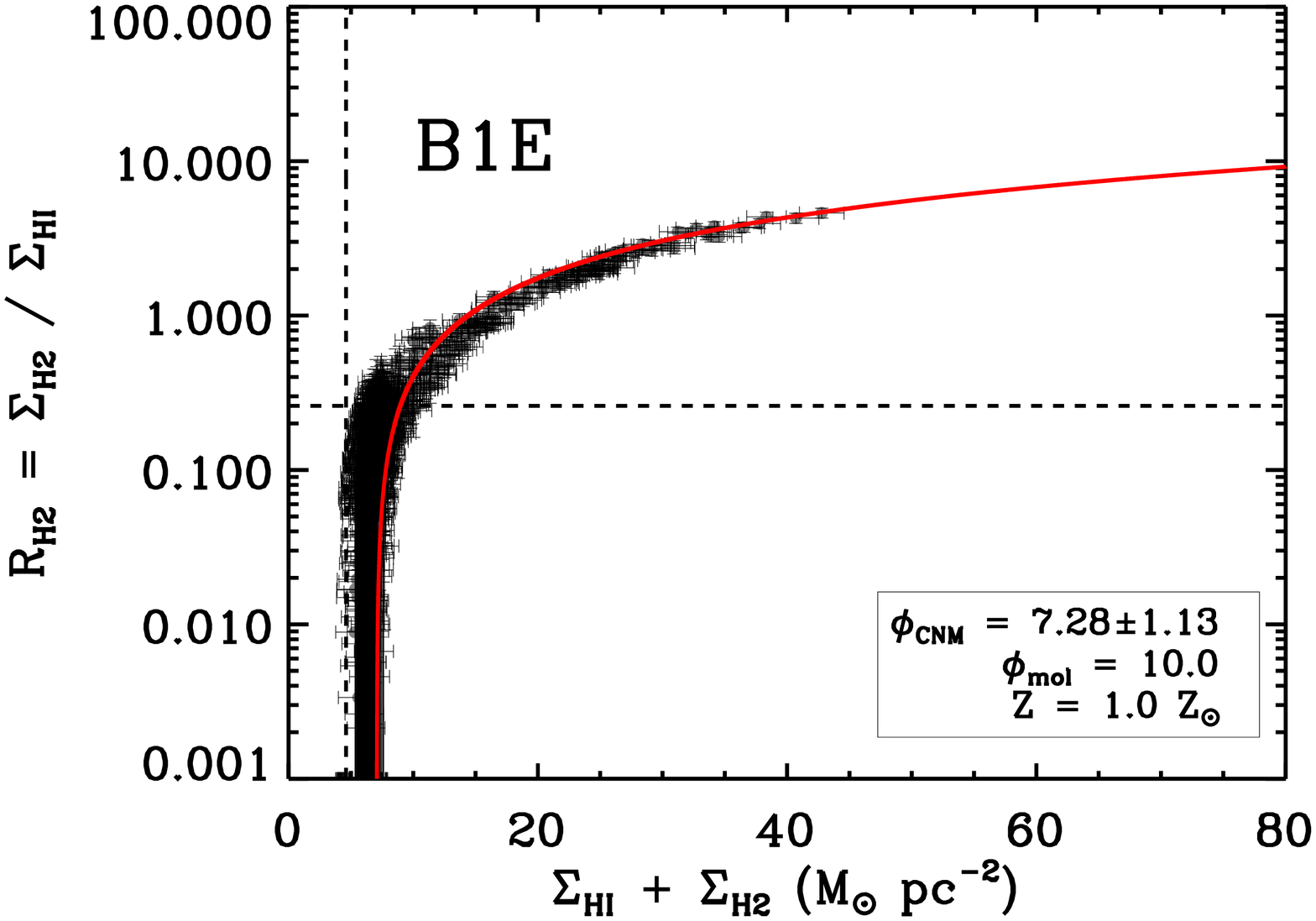}
\includegraphics[scale=0.33]{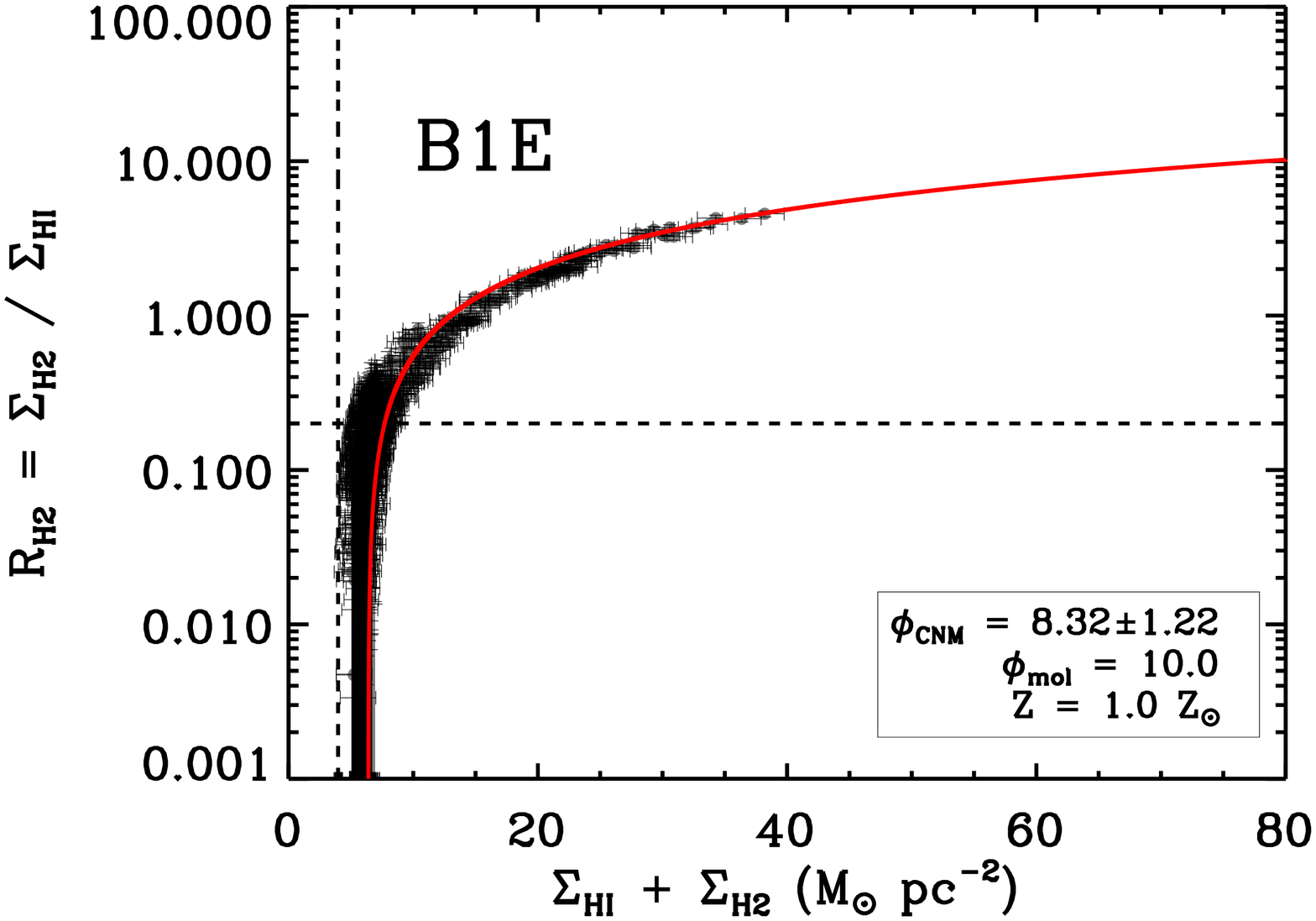}
\caption{\label{f:R-H2} (left) $R_{\rm H2}$ versus $\Sigma_{\rm HI}+\Sigma_{\rm H2}$ for IC348 and B1E (this study). 
All finite pixels in the rectangular boxes shown in Figure \ref{f:new-H2} are used for plotting. 
The median 3$\sigma$ values of $R_{\rm H2}$ and $\Sigma_{\rm HI}+\Sigma_{\rm H2}$ are indicated as the black dashed lines, 
while the best-fit curves determined in Section \ref{s:KMT09} are overlaid in red.
The best-fit parameters are shown in the bottom right corner of each plot. 
(right) Plots from \citet{Lee12}.} 
\end{figure*}

For the uncertainty in $N$(HI), we combine two terms in quadrature: 
the error from using a fixed velocity width ($\sigma_{\rm HI,1}$), 
and the error from the correction for high optical depth ($\sigma_{\rm HI,2}$).
To estimate $\sigma_{\rm HI,1}$, we produce 1000 $N$(HI) images using 1000 velocity widths 
randomly drawn from a Gaussian distribution that peaks at 20 km s$^{-1}$ with 1$\sigma$ of 4 km s$^{-1}$. 
The standard deviation of the simulated $N$(HI) is then computed for $\sigma_{\rm HI,1}$. 
This $\sigma_{\rm HI,1}$ is what \citet{Lee12} used as their uncertainty in $N$(HI). 
Similarly, to derive $\sigma_{\rm HI,2}$, we generate 1000 $N$(HI) images
by applying the correction to the $N$(HI) image from \citet{Lee12} 
using 1000 combinations of $a$ and $b$ in Equation (\ref{eq:f-carl}). 
These $a$ and $b$ values are again drawn from Gaussian distributions 
whose peaks and widths correspond to the fitted $a$ and $b$ values and their uncertainties. 
We find that the median 1$\sigma$ of $N$(HI) in our study is $\sim$8.1 $\times$ 10$^{19}$ cm$^{-2}$. 

For the uncertainty in $A_{V}$, we repeat the exercise done by \citet{Lee12}: 
deriving a number of $A_{V}$ images by changing input conditions 
(1$\sigma$ noises of the IRIS 60/100 $\mu$m images, $\beta$, zero point calibration for $\tau_{100}$), 
and estimating the minimum and maximum $A_{V}$ values for each pixel. 
In this exercise, we find that the contribution from VSGs to $I_{60}$ varies from 76\% to 88\%, 
while $X$ varies from 655 to 855.

Finally, we propagate the uncertainty in $N$(HI) and the minimum/maximum $A_{V}$ values through a Monte Carlo simulation 
in order to produce 1000 $N$(H$_{2}$) images.  
The distribution of the simulated $N$(H$_{2}$) is then used to estimate the uncertainty in $N$(H$_{2}$) on a pixel-by-pixel basis. 
We find that the median 1$\sigma$ of $N$(H$_{2}$) in our study is $\sim$4.8 $\times$ 10$^{19}$ cm$^{-2}$. 

\subsection{$R_{\rm H2}$ versus $\Sigma_{\rm HI}+\Sigma_{\rm H2}$ and 
$\Sigma_{\rm HI}$ versus $\Sigma_{\rm HI}+\Sigma_{\rm H2}$}
\label{s:R-H2-HI} 
From the rederived $N$(HI) and $N$(H$_{2}$) images,  
we estimate $\Sigma_{\rm HI}$ and $\Sigma_{\rm H2}$ by 
\begin{equation*}
\Sigma_{\rm HI}~(\rm M_{\odot}~pc^{-2}) = 
\frac{\textit{N}(HI)~(cm^{-2})}{1.25 \times 10^{20}}
\end{equation*}
\begin{equation}
\Sigma_{\rm H2}~(\rm M_{\odot}~pc^{-2}) = 
\frac{\textit{N}(H_{2})~(cm^{-2})}{6.25 \times 10^{19}} .
\end{equation}

\noindent We find that $\Sigma_{\rm HI}$ varies by only a factor of $\sim$2.6 
from $\sim$4.8 M$_{\odot}$ pc$^{-2}$ to $\sim$12.7 M$_{\odot}$ pc$^{-2}$. 
On the other hand, $\Sigma_{\rm H2}$ ranges from $-$2.4 M$_{\odot}$ pc$^{-2}$ to 81.8 M$_{\odot}$ pc$^{-2}$, 
although $\sim$98\% of the pixels have $\Sigma_{\rm H2}$ $<$ 15 M$_{\odot}$ pc$^{-2}$. 
As a result, $\Sigma_{\rm HI}+\Sigma_{\rm H2}$ has a small dynamic range of $\sim$8--30 M$_{\odot}$ pc$^{-2}$ 
across most of the cloud.

To compare with the KMT09 predictions aiming at revisiting the HI saturation in Perseus,  
we focus on the individual dark (B5, B1E, and B1) and star-forming (IC348 and NGC1333) regions. 
The boundaries of each region were determined based on the $^{13}$CO emission (Section 5 of \citeauthor{Lee12} 2012 for details),  
and are shown as the red rectangular boxes in Figure \ref{f:new-H2}. 
Using all finite pixels in the rectangular boxes, 
we plot $\Sigma_{\rm HI}$ and $R_{\rm H2}$ as a function of $\Sigma_{\rm HI}+\Sigma_{\rm H2}$ for each region, 
and present the results for IC348 and B1E in Figures \ref{f:HI} and \ref{f:R-H2}.
Additionally, we show the same plots from \citet{Lee12} for comparison. 
Note that both this study and \cite{Lee12} include negative $\Sigma_{\rm H2}$ values  
by using all finite pixels in the rectangular boxes. 
Almost all ($\sim$90\%) of these negative $\Sigma_{\rm H2}$ values fluctuate around zero within uncertainties. 

\subsection{Comparison to the KMT09 Predictions}
\label{s:KMT09}
As in \citet{Lee12}, the following KMT09 predictions are used to fit 
the observed $R_{\rm H2}$ vs $\Sigma_{\rm HI}+\Sigma_{\rm H2}$ profiles: 
\begin{equation}
R_{\rm H2} = \frac{4 \tau_{\rm c}}{3 \psi} \left[1 + \frac{0.8 \psi \phi_{\rm mol}}{4 \tau_{\rm c} + 
               3(\phi_{\rm mol} - 1) \psi}\right] - 1  
\end{equation}
\noindent where 
\begin{equation}
\tau_{\rm c} = \frac{3}{4} \left(\frac{\Sigma_{\rm comp} \sigma_{\rm d}}{\mu_{\rm H}}\right),
\end{equation}
\begin{equation} 
\psi = \chi \frac{2.5 + \chi}{2.5 + \chi e},
\end{equation}
\noindent and 
\begin{equation}
\chi = 2.3 ~ \frac{1 + 3.1 Z'^{0.365}}{\phi_{\rm CNM}}. 
\end{equation}

\noindent Here $\tau_{\rm c}$ is the dust optical depth a spherical cloud would have 
if its HI and H$_{2}$ are uniformly mixed,
and $\chi$ is the ratio of the rate at which LW photons are absorbed by dust grains 
to the rate at which they are absorbed by H$_{2}$. 
In addition, $\Sigma_{\rm comp}$ is the total gas column density, 
$\sigma_{\rm d}$ is the dust absorption cross section per hydrogen nucleus in the LW band, 
$\mu_{\rm H}$ is the mean mass per hydrogen nucleus, 
$Z'$ is the metallicity normalized to the value in the solar neighborhood, 
$\phi_{\rm CNM}$ is the ratio of the CNM density to the minimum CNM density 
at which the CNM can be in pressure balance with the WNM, 
and finally $\phi_{\rm mol}$ is the ratio of the H$_{2}$ density to the CNM density.
We refer to Section 6 of \citet{Lee12} for a detailed summary of the KMT09 model. 

As Equations (17)--(20) suggest, $R_{\rm H2}$ in the KMT09 model is simply 
a function of total gas column density, metallicity, $\phi_{\rm CNM}$, and $\phi_{\rm mol}$, 
and is independent of the strength of the radiation field in which the cloud is embedded. 
Following \citet{Lee12}, 
we adopt $Z'$ = 1 and $\phi_{\rm mol}$ = 10 (fiducial value by KMT09)\footnote{We note that 
$R_{\rm H2}$ is not sensitive to $\phi_{\rm mol}$. 
For example, with our median $\phi_{\rm CNM}$ value of $\sim$7 (Table 1), 
$R_{\rm H2}$ at $\Sigma_{\rm HI}+\Sigma_{\rm H2}$ = 100 M$_{\odot}$ pc$^{-2}$ varies by only a factor of $\sim$1.1 
for $\phi_{\rm mol}$ = 10--50.}, and constrain $\phi_{\rm CNM}$ 
by finding the best-fit model for the observed $R_{\rm H2}$ vs $\Sigma_{\rm HI}+\Sigma_{\rm H2}$. 
For this purpose, we perform Monte Carlo simulations 
where the uncertainties in $R_{\rm H2}$ and $\Sigma_{\rm HI}+\Sigma_{\rm H2}$ 
are taken into account for model fitting. 
In these simulations, we add random offsets to $R_{\rm H2}$ and $\Sigma_{\rm HI}+\Sigma_{\rm H2}$ based on their uncertainties, 
and determine the best-fit curve by setting $Z'$ = 1 and $\phi_{\rm mol}$ = 10 
and finding $\phi_{\rm CNM}$ that results in the minimum sum of squared residuals. 
We repeat this process 1000 times, and estimate the best-fit $\phi_{\rm CNM}$ 
by calculating the median $\phi_{\rm CNM}$ among the simulated $\phi_{\rm CNM}$.
The derived $\phi_{\rm CNM}$ for each region is summarized in Table \ref{t:fitting}, 
and the best-fit curves are shown in red in Figures \ref{f:HI} and \ref{f:R-H2}. 

\begin{table}
\begin{center}
\caption{\label{t:fitting} Fitting Results for $R_{\rm H2}$ versus $\Sigma_{\rm H\textsc{i}}+\Sigma_{\rm H2}$}
\begin{tabular}{l c}\hline \hline
Region & Best-fit $\phi_{\rm CNM}$ \\ \hline
B5 & $8.75 \pm 1.35$ \\ 
IC348 & $7.40 \pm 0.94$ \\
B1E & $7.28 \pm 1.13$ \\
B1 & $6.93 \pm 1.00$ \\
NGC1333 & $5.28 \pm 0.85$ \\
\hline
\end{tabular}
\end{center}
{Note. The uncertainty in $\phi_{\rm CNM}$ is estimated from 
the distribution of the simulated 1000 $\phi_{\rm CNM}$ values.}
\end{table}

In the KMT09 model, $\chi$ measures the relative importance of dust shielding and H$_{2}$ self-shielding,  
and is predicted to be $\sim$1 for a wide range of galactic environments. 
In this case, a certain amount of $\Sigma_{\rm HI}$ is required to shield H$_{2}$ againt photodissociation, 
and H$_{2}$ forms out of HI once this minimum shielding column density is obtained. 
The KMT09 model predicts the minimum shielding column density of $\sim$10 M$_{\odot}$ pc$^{-2}$ for solar metallicity, 
and this is indeed consistent with what we observe in Perseus: 
$\Sigma_{\rm HI}$ saturates at $\sim$7--9 M$_{\odot}$ pc$^{-2}$ for all five regions. 
The level of the HI saturation changes between the regions though, 
from $\sim$7 M$_{\odot}$ pc$^{-2}$ for B5 to $\sim$9 M$_{\odot}$ pc$^{-2}$ for NGC1333. 
In comparison with \citet{Lee12}, our $\Sigma_{\rm HI}$ values are slightly higher 
due to the correction for high optical depth. 
This correction brings a closer agreement with the KMT09 model. 

The excellent agreement with the KMT09 model is also evident from 
$R_{\rm H2}$ vs $\Sigma_{\rm HI}+\Sigma_{\rm H2}$ in Figure \ref{f:R-H2}. 
For all five regions, we find that $R_{\rm H2}$ steeply rises at small $\Sigma_{\rm HI}+\Sigma_{\rm H2}$, 
turns over at $R_{\rm H2}$ $\sim$ 1, and then slowly increases at large $\Sigma_{\rm HI}+\Sigma_{\rm H2}$. 
In fact, this common trend on a log-linear scale
is entirely driven by the almost constant $\Sigma_{\rm HI}$, 
and therefore it is not surprising to find such a good agreement with the KMT09 model 
where the $\Sigma_{\rm HI}$ saturation is predicted.
By fitting the KMT09 predictions to the observed $R_{\rm H2}$ vs $\Sigma_{\rm HI}+\Sigma_{\rm H2}$ profiles, 
we constrain $\phi_{\rm CNM}$ $\sim$ 5--9,  
which is consistent with what \cite{Lee12} estimated within uncertainties. 
In the KMT09 model, $\phi_{\rm CNM}$ determines the CNM density by 
$n_{\rm CNM} = \phi_{\rm CNM}n_{\rm min}$  
where $n_{\rm min}$ is the minimum CNM density at which the CNM can be in pressure balance with the WNM, 
and $\phi_{\rm CNM}$ = 5--9 translates into $T_{\rm CNM}$ = 60--80 K (Equation 19 of KMT09). 
This $T_{\rm CNM}$ range is consistent with 
what we observationally constrained for the HI environment of Perseus via the HI absorption measurements (Paper I): 
$T_{\rm CNM}$ mostly ranges for $\sim$10--200 K, and its distribution peaks at $\sim$50 K. 
Finally, we note that $\phi_{\rm CNM}$ systematically decreases toward the southwest of Perseus, 
reflecting the observed region-to-region variations in $\Sigma_{\rm HI}$.
The difference in $\phi_{\rm CNM}$, however, is not significant, 
and this suggests similar $\chi$ values for all dark and star-forming regions (Equation 20).  
We indeed find $\chi$ = $\sim$1.1--1.8, in agreement with the KMT09 prediction of   
comparable dust-shielding and H$_{2}$ self-shielding for H$_{2}$ formation. 

\section{Optically Thick HI: Alternative to the ``CO-dark'' Gas?}
\label{s:CO-dark}

Recently, Fukui et al. (2014,2015) suggested that the ``CO-dark'' gas in the Galaxy, 
referring to the interstellar gas undetectable either in the 21-cm HI and 2.6-mm CO emission (e.g., \citeauthor{Bolatto13} 2013), 
could be dominated by the optically thick HI.
As our HI absorption measurements provide an independent estimate of the optical depth, 
we can test the validity of their claim against Perseus on sub-pc scales. 
To do so, we utilize our old and new $N$(H$_{2}$) images 
(before and after the correction for high optical depth), 
as well as the CfA $I_{\rm CO}$ data. 

First, we examine how the ``CO-dark'' gas and the optically thick HI gas are spatially distributed. 
In order to identify the ``CO-dark'' gas, 
we use the 3$\sigma$ contours of the old $N$(H$_{2}$) and CfA $I_{\rm CO}$ images, 
following the definition by \citet{Lee12}. 
These contours are overlaid on our correction factor image (Figure \ref{f:corr-fact}), 
and show that the relative distribution of H$_{2}$ (or simply the gas not probed by the HI emission)
and CO changes across the cloud.   
For example, H$_{2}$ and CO trace each other in the southwest,  
while H$_{2}$ is more extended than CO elsewhere.  
\cite{Lee12} compared H$_{2}$ and CO radial profiles for Perseus, 
and estimated that H$_{2}$ is on average $\sim$1.4 times more extended than CO, 
suggesting a substantial amount of the ``CO-dark'' gas. 
We then find that the distributions of the ``CO-dark'' gas (traced by the difference between the H$_{2}$ and CO 3$\sigma$ contours) 
and the optically thick HI gas (traced by the correction factor) generally disagree with each other.  
For example, the region around B5 where the discrepancy between the H$_{2}$ and CO distributions is greatest 
shows relatively low correction factors.
On the other hand, the regions with high correction factors at (R.A.,decl.) $\sim$ (3$^{\rm h}$33$^{\rm m}$,$+$34$^{\circ}$30$'$)
and $\sim$ (3$^{\rm h}$23$^{\rm m}$,$+$29$^{\circ}$) do not have a large amount of the ``CO-dark'' gas. 

\begin{figure} 
\centering 
\includegraphics[scale=0.48]{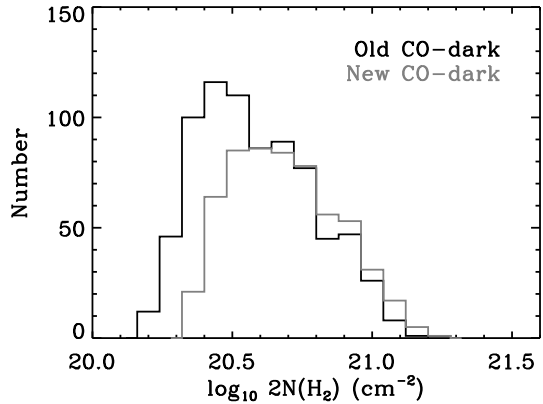}
\caption{\label{f:CO-dark} Histograms of the ``CO-dark'' gas column density. 
The ``CO-dark'' gas from the old $N$(H$_{2}$) image is in black, 
while that from the new $N$(H$_{2}$) image is in gray. 
Both histograms are constructed using the data smoothed to 8.4$'$ resolution.}
\end{figure}

While our previous evaluation was based on the visual examination of the 3$\sigma$ contours, 
here we more rigorously investigate whether or not the optically thick HI gas can explain the ``CO-dark'' gas 
by comparing the old and new $N$(H$_{2}$) images. 
We first smooth the old and new $N$(H$_{2}$) images to 8.4$'$ resolution, 
as well as their uncertainties, 
to match the resolution of the CfA $I_{\rm CO}$ image. 
We then find the ``CO-dark'' gas from the smoothed $N$(H$_{2}$) images
by utilizing the 3$\sigma$ contours of $N$(H$_{2}$) and $I_{\rm CO}$. 
Essentially, a pixel is classified as the ``CO-dark'' gas 
when $N$(H$_{2}$) is above the 3$\sigma$ level, 
but $I_{\rm CO}$ is less than the 3$\sigma$ noise. 
For the selected pixels, we calculate the ``CO-dark'' gas column density, 2$N$(H$_{2}$),
and present two histograms in Figure \ref{f:CO-dark}. 
The ``CO-dark'' gas from the old $N$(H$_{2}$) is in black, 
while that from the new $N$(H$_{2}$) is in gray.
Figure \ref{f:CO-dark} shows that the two histograms are comparable 
regarding their minimum, maximum, and median values (different by less than a factor of 2),  
although the gray histogram has a smaller number of pixels due to the larger uncertainty of the new $N$(H$_{2}$) image. 
Given that the optically thick HI gas was already taken into consideration in the derivation of the new $N$(H$_{2}$) image, 
the comparable black and gray histograms suggest that the increased column density 
due to the optically thick HI gas is small (up to $\sim$2 $\times$ 10$^{20}$ cm$^{-2}$),
and the ``CO-dark'' gas still exists in Perseus. 
In terms of mass, the additional contribution from the optically thick HI only accounts for $\sim$20\% of the observed ``CO-dark'' gas. 

While our results are in contrast with Fukui et al. (2014,2015) 
who found that the optically thick HI adds the column density of $\sim$10$^{20}$--10$^{22}$ cm$^{-2}$
and possibly explains the ``CO-dark'' gas in the Galaxy, 
there are multiple factors that could affect the comparison, e.g.,  
spatial coverage ($\sim$500 deg$^{2}$ for Perseus vs whole Galactic sky at $|b|$ $>$ 15$^{\circ}$) 
and method for deriving $N$(H$_{2}$) (IRIS/2MASS vs \textit{Planck}). 
In particular, we note that this study and Fukui et al. (2014,2015) probe very different scales: 
our results are based on pencil-beam HI absorption measurements on 3.5$'$ scales, 
while Fukui et al. (2014,2015) estimated the correction factor on 33$'$ scales. 
If the CNM is highly structured with a low filling factor, 
this could affect the estimate of the correction factor in both studies. 
In the future, it will be important to compare the results from Fukui et al. (2014,2015) 
with a large sample of HI absorption measurements as well as numerical simulations 
(e.g., \citeauthor{Audit05} 2005; \citeauthor{Kim14} 2014)
to investigate how the derivation of the correction factor depends on different methodologies 
and CNM properties.

\section{Summary}
\label{s:summary}

In this paper, we investigate the impact of high optical depth 
on the HI column density distribution across the Perseus molecular cloud. 
We use Arecibo HI emission and absorption measurements obtained toward 26 background sources (Paper I)
in order to derive the properties of CNM and WNM components 
along each line of sight via the Gaussian decomposition approach (HT03a). 
The derived properties are then used to estimate the correction factor for high optical depth, 
and the correction is applied to the HI column density image computed in the optically thin approximation.

To revisit the HI saturation in Perseus observed by \cite{Lee12},  
we rederive the H$_{2}$ column density image by adopting the same methodology as \cite{Lee12}, 
but using the corrected HI column density image. 
The final HI and H$_{2}$ column density images at $\sim$0.4 pc resolution are then compared with the KMT09 predictions. 
Finally, we investigate if the observed ``CO-dark'' gas in Perseus is dominated by the optically thick HI gas.  

We summarize our main results as follows. 

\begin{enumerate}[topsep=0pt,itemsep=-1ex]
 
\item We estimate the correction factor for high optical depth ($f$), 
which is defined as the ratio of the true total HI column density ($N_{\rm tot}$) 
to the HI column density derived in the optically thin approximation ($N_{\rm exp}$), 
and express it as a function of $N_{\rm exp}$: 
$f$ = log$_{10}$($N_{\rm exp}$/10$^{20}$)(0.32 $\pm$ 0.06) + (0.81 $\pm$ 0.05). 
We use two different methods, Gaussian decomposition and isothermal approximation methods, 
and find that they are consistent within uncertainties. 
This is likely due to the relatively low optical depth and 
insignificant contribution from the diffuse radio continuum emission for Perseus.   

\item We estimate that the correction factor in/around Perseus is small (up to $\sim$1.2), 
and the total HI mass increases by only $\sim$10\% from $\sim$2.3 $\times$ 10$^{4}$ M$_{\odot}$ 
to $\sim$2.5 $\times$ 10$^{4}$ M$_{\odot}$ due to the inclusion of the optically thick HI gas. 

\item The H$_{2}$ column density image rederived using the corrected HI column density image 
is comparable to the original one by \cite{Lee12}, 
confirming the minor correction for high optical depth. 

\item For individual dark and star-forming regions in Perseus (B5, B1E, B1, IC348, and NGC1333), 
the HI surface density is relatively uniform with $\sim$7--9 M$_{\odot}$ pc$^{-2}$. 
This is slightly higher than what \cite{Lee12} found 
due to the correction for high optical depth. 
The correction brings a closer agreement with the KMT09 model  
where the minimum HI surface density of $\sim$10 M$_{\odot}$ pc$^{-2}$ is predicted 
for shielding H$_{2}$ against photodissociation in the solar metallicity environment. 

\item Driven by the uniform $\Sigma_{\rm HI}$ $\sim$ 7--9 M$_{\odot}$ pc$^{-2}$, 
$R_{\rm H2}$ vs $\Sigma_{\rm HI} + \Sigma_{\rm H2}$ on a log-linear scale 
shows remarkably consistent results for all dark and star-forming regions: 
$R_{\rm H2}$ sharply rises at small $\Sigma_{\rm HI} + \Sigma_{\rm H2}$, 
and then gradually increases toward large $\Sigma_{\rm HI} + \Sigma_{\rm H2}$ 
with the transition at $R_{\rm H2}$ $\sim$ 1. 

\item The mass increase due to the optically thick HI only accounts for $\sim$20\% of the observed ``CO-dark'' gas in Perseus. 
The spatial distributions of the ``CO-dark'' gas and the optically thick HI gas do not generally coincide with each other,  
and the ``CO-dark'' gas still exists even after the optically thick HI is considered in the derivation of H$_{2}$. 

\end{enumerate} 

Our study is one of the first attempts to examine the properties of the cold and warm HI in molecular cloud environments
and their relation with the HI-to-H$_{2}$ transition. 
While HI envelopes surrounding molecular clouds have been frequently found  
(e.g., \citeauthor{Knapp74} 1974; \citeauthor{Wannier83} 1983; \citeauthor{Elmegreen87} 1987; 
\citeauthor{Andersson91} 1991; \citeauthor{Rogers95} 1995; \citeauthor{Williams96} 1996; 
\citeauthor{Imara11} 2011; \citeauthor{Lee12} 2012; \citeauthor{Lee14} 2014; \citeauthor{Motte14} 2014), 
HI has traditionally been considered less important for the formation and evolution of molecular clouds. 
The excellent agreement between our observations of Perseus on sub-pc scales and the KMT09 model, on the other hand,  
suggests the significance of HI as one of the key ingredients for the HI-to-H$_{2}$ transition and consequently for star formation.

While our data agree with the KMT09 model, 
further theoretical developments are required.  
For example, the KMT09 model considers only the CNM as a source of shielding against H$_{2}$ photodissociaiton.
Our HI emission and absorption measurements, however, 
show that the CNM and the WNM have comparable column density contributions for Perseus (Paper I).
Clearly, the WNM needs to be taken into consideration in the models of H$_{2}$ formation (e.g., Bialy et al. in prep). 
In addition, the KMT09 model ignores the effect of internal radiation field. 
This is valid for Perseus, as there are no early-type stars producing a significant amount of dissociating radiation (\citeauthor{Lee12} 2012).
However, the role of internal radiation field would be important for massive molecular clouds containing a large number of OB stars. 
The discrepancy with the KMT09 model recently found for the W43 molecular cloud complex in the Galactic plane (\citeauthor{Motte14} 2014) 
and the low-metallicity SMC (\citeauthor{Welty12} 2012) suggests that 
additional physical ingredients (e.g., better description of H$_{2}$ formation and photodissociation in the low-metallicity ISM 
and strong shear/turbulence at galactic bars) would be necessary for extreme environments.
Finally, vertical thermal and dynamical equilibrium in a galactic disk is another important aspect, 
as recently explored by \cite{Kim14}. 
Future comparisons between theoretical models and HI emission/absorption observations of molecular clouds in a wide range of ISM environments
will be important for a deep understanding of HI properties and their role in the formation and evolution of molecular clouds.

\acknowledgements
We thank the anonymous referee for suggestions that improved this work. 
We also thank Shmuel Bialy, John Dickey, Yasuo Fukui, Mark Krumholz, Vianney Lebouteiller, Franck Le Petit, Harvey Liszt, 
Suzanne Madden, Evelyne Roueff, and Amiel Sternberg for stimulating discussion, 
and telescope operators at the Arecibo Observatory for their help in conducting our HI observations. 
M.-Y.L acknowledges supports from the DIM ACAV of the Region Ile de France, 
and S.S acknowledges supports from the NSF Early Career Development (CAREER) Award AST-1056780. 
We also acknowledge the NSF REU grant AST-1004881, which funded summer research of Jesse Miller. 
For this work, we have made use of the KARMA visualization software (Gooch 1996) and NASA's Astrophysics Data System (ADS). 

\appendix
\section{Comparison between Method 1 and Method 2} 
In order to investigate why the Gaussian decomposition and isothermal methods 
result in comparable correction factors despite of their apparent differences (Sections \ref{s:method1} and \ref{s:method2}), 
we start with Equation (\ref{eq:N-true-JD}), and compare it with Equation (\ref{eq:N-true}):  
\begin{equation}
\label{eq:A1} 
N_{\textrm{tot}}~(\textrm{cm}^{-2}) = C_{0} \int f_{\textrm{chan}}(v)T_{\textrm{exp}}(v)dv
\end{equation}
\noindent 
where $C_0 = 1.823 \times 10^{18}$ cm$^{-2}$/(K km s$^{-1}$). 
Combined with Equation (\ref{eq:f-chan3}), Equation (\ref{eq:A1}) then becomes 
\begin{equation} 
\label{eq:A2} 
\begin{split}
N_{\textrm{tot}}~(\textrm{cm}^{-2}) & = C_{0} \int \frac{T_{\textrm{s}}(v)}{T_{\textrm{s}}(v) - T_{\textrm{sky}}} 
                                                 \frac{\tau(v)}{1 - e^{-\tau(v)}} T_{\textrm{exp}}(v)dv \\
                                    & \sim C_{0} \int \frac{\tau(v)}{1 - e^{-\tau(v)}} T_{\textrm{exp}}(v)dv.
\end{split}
\end{equation}
\noindent 
The approximation in Equation (\ref{eq:A2}) can be made 
due to the small $T_{\rm s}(v)/[T_{\rm s}(v) - T_{\rm sky}]$ of $\sim$1.0--1.2 (Section \ref{s:method2}). 

While an optical depth spectrum is primarily determined by the CNM, 
an ``expected'' emission spectrum has contributions from both the CNM and the WNM. 
Hence, $T_{\rm exp}(v)$ can be expressed as (Equations 2, 3, and 4 of Paper I):
\begin{equation}
\label{eq:A3} 
T_{\textrm{exp}}(v) = T_{\textrm{B,CNM}}(v) + T_{\textrm{B,WNM}}(v)
\end{equation}
\noindent 
with 
\begin{gather*}
\label{eq:A3-sub}
T_{\textrm{B,CNM}}(v) = \sum_{0}^{N-1} T_{\textrm{s},n} (1 - e^{-\tau_{n}(v)})e^{-\sum_{0}^{M-1} \tau_{m}(v)} \\ 
T_{\textrm{B,WNM}}(v) = \sum_{0}^{K-1} [F_{k} + (1 - F_{k})e^{-\tau(v)}]T_{0,k}e^{-[(v - v_{0,k})/\delta v_{k}]^{2}}  
\end{gather*}  
\noindent 
where 
$T_{\rm B,CNM}(v)$ is the emission from $N$ CNM components, 
$T_{\rm B, WNM}(v)$ is the emission from $K$ WNM components, 
the components with subscript $n$ refer to the CNM, 
the components with subscript $k$ refer to the WNM, 
$\tau_{m}$ is the optical depth of $m$ CNM component that lies in front of $n$ cloud, 
$F_{k}$ is the fraction of $k$ WNM component that lies in front of all CNM clouds, 
$T_{0}$ is the peak brightness temperature, 
$v_{0}$ is the central velocity, 
and finally $\delta v$ is the 1/$e$ width of the component. 
Equation (\ref{eq:A2}) then becomes 
\begin{equation}
\label{eq:A4}
N_{\textrm{tot}}~(\textrm{cm}^{-2}) \sim C_{0} \int \frac{\tau(v)}{1 - e^{-\tau(v)}} 
                                     \left\{\sum_{0}^{N-1} T_{\textrm{s},n} (1 - e^{-\tau_{n}(v)})e^{-\sum_{0}^{M-1} \tau_{m}(v)} + 
                                            \sum_{0}^{K-1} [F_{k} + (1 - F_{k})e^{-\tau(v)}]T_{0,k}e^{-[(v - v_{0,k})/\delta v_{k}]^{2}}\right\}dv. 
\end{equation}

We examine Equation (\ref{eq:A4}) under two extreme circumstances: 
(I) $F_{k}$ = 0 and $\tau \ll 1$ and (II) $F_{k}$ = 1 and $\tau \ll 1$.  
In both cases, the first part of Equation (\ref{eq:A4}) becomes 
\begin{equation}
\label{eq:A5} 
\begin{split} 
A & = C_{0} \int \frac{\tau(v)}{1 - e^{-\tau(v)}}\left\{\sum_{0}^{N-1} T_{\textrm{s},n} (1 - e^{-\tau_{n}(v)}) e^{-\sum_{0}^{M-1} \tau_{m}(v)}\right\}dv \\ 
  & \sim C_{0} \int \frac{\tau(v)}{1 - e^{-\tau(v)}}\left\{\sum_{0}^{N-1} T_{\textrm{s},n} (1 - e^{-\tau_{n}(v)})\right\}dv \\ 
  & \sim C_{0} \int \frac{\tau(v)}{1 - (1 - \tau(v))}\left\{\sum_{0}^{N-1} T_{\textrm{s},n} (1 - (1 - \tau_{n}(v)))\right\}dv \\
  & = C_{0} \int \sum_{0}^{N-1} T_{\textrm{s},n} \tau_{n}(v)dv \\ 
  & = C_{0} \int \sum_{0}^{N-1} T_{\textrm{s},n} \tau_{0,n} e^{-[(v - v_{0,n})/\delta v_{n}]^2}dv. \\ 
\end{split}
\end{equation}
\noindent 
On the other hand, the second part of Equation (\ref{eq:A4}) becomes \\
\noindent 
(I) $F_{k}$ = 0 (all WNM components are behind the CNM) and $\tau \ll 1$: \\
\begin{equation} 
\label{eq:A6} 
\begin{split}
B & = C_{0} \int \frac{\tau(v)}{1 - e^{-\tau(v)}}\left\{\sum_{0}^{K-1} [F_{k} + (1 - F_{k})e^{-\tau(v)}]T_{0,k}e^{-[(v - v_{0,k})/\delta v_{k}]^{2}}\right\}dv \\           & = C_{0} \int \frac{\tau(v)e^{-\tau(v)}}{1 - e^{-\tau(v)}}\left\{\sum_{0}^{K-1} T_{0,k}e^{-[(v - v_{0,k})/\delta v_{k}]^{2}}\right\}dv \\
  & \sim C_{0} \int \frac{\tau(v)}{1 - (1 - \tau(v))}\left\{\sum_{0}^{K-1} T_{0,k}e^{-[(v - v_{0,k})/\delta v_{k}]^{2}}\right\}dv \\
  & = C_{0} \int \sum_{0}^{K-1} T_{0,k}e^{-[(v - v_{0,k})/\delta v_{k}]^{2}}dv. \\ 
\end{split}
\end{equation}
(II) $F_{k}$ = 1 (all WNM components are in front of the CNM) and $\tau \ll 1$: \\
\begin{equation}
\label{eq:A7} 
\begin{split} 
B & = C_{0} \int \frac{\tau(v)}{1 - e^{-\tau(v)}}\left\{\sum_{0}^{K-1} [F_{k} + (1 - F_{k})e^{-\tau(v)}]T_{0,k}e^{-[(v - v_{0,k})/\delta v_{k}]^{2}}\right\}dv \\           & = C_{0} \int \frac{\tau(v)}{1 - e^{-\tau(v)}}\left\{\sum_{0}^{K-1} T_{0,k}e^{-[(v - v_{0,k})/\delta v_{k}]^{2}}\right\}dv \\
  & \sim C_{0} \int \frac{\tau(v)}{1 - (1 - \tau(v))}\left\{\sum_{0}^{K-1} T_{0,k}e^{-[(v - v_{0,k})/\delta v_{k}]^{2}}\right\}dv \\
  & = C_{0} \int \sum_{0}^{K-1} T_{0,k}e^{-[(v - v_{0,k})/\delta v_{k}]^{2}}dv. \\
\end{split}
\end{equation}
\noindent 
As a result, we have 
\begin{equation} 
\label{eq:A8} 
\begin{split}
N_{\textrm{tot}}~(\textrm{cm}^{-2}) & = A + B \\ 
                                    & \sim C_{0} \int \left\{\sum_{0}^{N-1} T_{\textrm{s},n} \tau_{0,n} e^{-[(v-v_{0,n}) / \delta v_{n}]^2} + 
                                                             \sum_{0}^{K-1} T_{0,k} e^{-[(v-v_{0,k})/\delta v_{k}]^2}\right\}dv,
\end{split}
\end{equation}
\noindent 
which is essentially Equation (\ref{eq:N-true}). 

These examples of the two extreme cases suggest that the Gaussian decomposition and isothermal methods 
would result in comparable correction factors ($f = N_{\rm tot}/N_{\rm exp}$)   
regardless of $F_{k}$ when $\tau \ll 1$ and $T_{\rm sky} \ll T_{\rm s}$. 

\section{Correction Factor in Heiles \& Troland (2003a,b)} 

\begin{figure}
\centering
\includegraphics[trim=0cm 0cm 15cm 0cm,clip=true,scale=0.55]{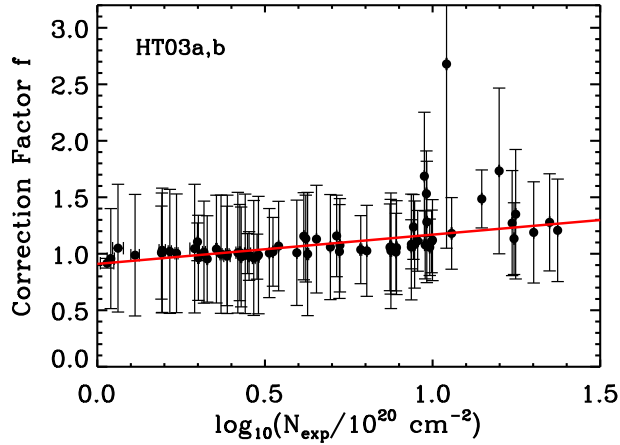}
\caption{\label{f:fcarl-HT03} $f$ = $N_{\rm tot}/N_{\rm exp}$ as a function of log$_{10}(N_{\rm exp}/10^{20})$ for 68 sources in HT03a,b. 
11 lines of sight probing the Galactic plane ($|b|$ $<$ 10$^{\circ}$) are excluded from the analysis to make a comparison with our study of Perseus.  
Both the true and optically thin HI column densities ($N_{\rm tot}$ and $N_{\rm exp}$) are calculated using the published full line of sight information,
and the linear fit to all 68 data points is overlaid in red: $f$ = log$_{10}$($N_{\rm exp}$/$10^{20})(0.26\pm0.08) + (0.91\pm0.04$).}
\end{figure}

To examine the validity of the linear relation we assume for the correction factor and the optically thin HI column density
(Sections \ref{s:method1} and \ref{s:method2}), 
we produce the same plot as Figure \ref{f:fcarl} (left) for HT03a,b. 
Among the total 79 sources in HT03a,b, 11 sources with $|b|$ $<$ 10$^{\circ}$ are excluded for a fair comparison with our study.  
The published ``expected'' emission spectra and total HI column densities are used to estimate 
$f$ and log$_{10}(N_{\rm exp}/10^{20})$ in the same way as we do for our 26 sources,  
and the results are presented in Figure \ref{f:fcarl-HT03}.

We find that Figure \ref{f:fcarl-HT03} is very similar with Figures \ref{f:fcarl} (left) and \ref{f:fjd} (left),   
suggesting that our assumption of the linear relation between $f$ and log$_{10}(N_{\rm exp}/10^{20})$ for Perseus is reasonable. 
Interestingly, however, a few sources with log$_{10}$($N_{\rm exp}$/$10^{20})$ $\gtrsim$ 1 show some deivation from the linear relation, 
which requires a further examination. 
See Section \ref{s:method2} for a more discussion.

\bibliographystyle{aa}
\bibliography{myref}

\end{document}